\documentclass[pra,twocolumn,aps,superscriptaddress]{revtex4-1}
\usepackage{amssymb}
\usepackage{amsmath}
\usepackage{graphicx}
\usepackage{color}
\usepackage{bm}
\usepackage[bookmarks=false,linkcolor=blue,urlcolor=blue,colorlinks,citecolor=blue]{hyperref}

\begin{document}
\author{Tommy Li}
\affiliation{Center for Quantum Devices and Station Q Copenhagen, Niels Bohr Institute,
University of Copenhagen, DK-2100 Copenhagen, Denmark}
\author{William A. Coish}
\affiliation{Center for Quantum Devices and Station Q Copenhagen, Niels Bohr Institute,
University of Copenhagen, DK-2100 Copenhagen, Denmark}
\affiliation{Department of Physics, McGill University, Montreal, Canada}
\affiliation{Quantum Information Science Program, Canadian Institute for Advanced Research, Toronto, Canada}
\affiliation{School of Physics, University of New South Wales, Sydney 2052, Australia}
\author{Michael Hell}
\affiliation{Center for Quantum Devices and Station Q Copenhagen, Niels Bohr Institute,
	University of Copenhagen, DK-2100 Copenhagen, Denmark}
\affiliation{Division of Solid State Physics and NanoLund, Lund University, Box 118, S-221 00 Lund, Sweden}
\author{Karsten Flensberg}
\affiliation{Center for Quantum Devices and Station Q Copenhagen, Niels Bohr Institute,
University of Copenhagen, DK-2100 Copenhagen, Denmark}
\author{Martin Leijnse}
\affiliation{Center for Quantum Devices and Station Q Copenhagen, Niels Bohr Institute,
University of Copenhagen, DK-2100 Copenhagen, Denmark}
\affiliation{Division of Solid State Physics and NanoLund, Lund University, Box 118, S-221 00 Lund, Sweden}

\title{Four-Majorana qubit with charge readout: dynamics and decoherence}

\begin{abstract}
  We present a theoretical analysis of a Majorana-based qubit consisting of two topological superconducting islands connected via a Josephson junction. The qubit is operated by electrostatic gates which control the coupling of two of the four Majorana zero modes. At the end of the operation, readout is performed in the charge basis. Even though the operations are not topologically protected, the proposed experiment can potentially shed light on the coherence of the parity degree of freedom in Majorana devices and serve as a first step towards topological Majorana qubits. We discuss in detail the charge-stability diagram and its use for characterizing the parameters of the devices, including the overlap of the Majorana edge states. We describe the multi-level spectral properties of the system and present a detailed study of its controlled coherent oscillations, as well as decoherence resulting from coupling to a non-Markovian environment. In particular, we study a gate-controlled protocol where conversion between Coulomb-blockade and transmon regimes generates coherent oscillations of the qubit state due to the overlap of Majorana modes. We show that, in addition to fluctuations of the Majorana coupling, considerable measurement errors may be accumulated during the conversion intervals when electrostatic fluctuations in the superconducting islands are present. These results are also relevant for several proposed implementations of topological qubits which rely on readout based on charge detection.
  \end{abstract}
\maketitle

\section{Introduction}

Majorana zero modes, hypothesized to occur in the vortices of two-dimensional (2D) $p$-wave superconductors~\cite{Read2000} and superfluids~\cite{Volovik1989}, have been recognized as a promising basis for fault-tolerant quantum computation~\cite{Kitaev2001, Nayak2008}. During the last decade, the interest in Majorana zero modes has increased~\cite{AliceaReview, LeijnseReview, BeenakkerReview} as a number of works showed that $p$-wave superconductivity can be engineered in, for example, topological insulators~\cite{Fu2008}, 2D electron gases in semiconductor heterostructures~\cite{Sau2010b, Alicea2010} and 1D semiconductor nanowires~\cite{Lutchyn2010,Oreg2010}. In the chain of developments towards this goal, significant recent evidence has indicated that these states have been observed by tunneling spectroscopy as the edge modes of semiconducting nanowires or two-dimensional electron gases proximitized by a bulk $s$-wave superconductor \cite{Das2012,Mourik2012,Deng2012,Churchill2013,Finck2013,Deng2016,Nichele2017,Zhang2018,Deng2018,Gul2018}.
\begin{figure}
\includegraphics[width=0.5\textwidth]{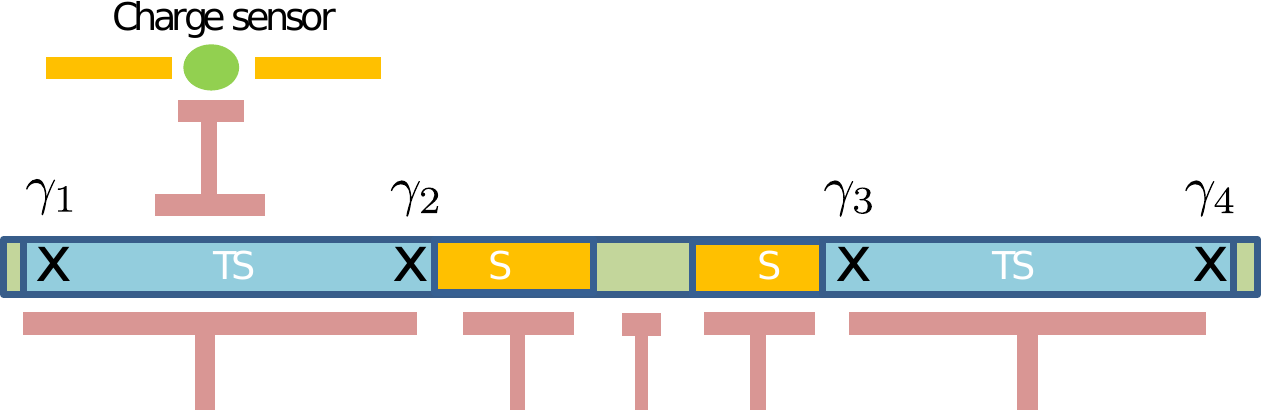}
\caption{\label{fig:device}(\emph{Color online}) Schematic of our system: the outer regions consist of two topological superconducting islands (blue) each hosting a pair of Majorana modes (black crosses) corresponding to the quasiparticle operators $\gamma_1, \gamma_2, \gamma_3, \gamma_4$. The gates shown below control the local potentials in the topological superconductors. The inner region consists of a gate-controlled tunnel junction (green) as well as two  trivial superconducting regions (orange). The two gates shown below the trivial superconductors are used to tune the Josephson coupling $E_J$. A charge sensor is used to perform a charge measurement on the left island.}
\end{figure}

The original conception of topological quantum computation relied on spatial manipulation of the zero modes in braiding operations \cite{Ivanov2001,Nayak2008,Alicea2011}. Several recent studies have proposed experimentally simpler schemes using Majorana islands controlled by electrostatic gates, whose operation is assisted by the charging energy of the islands, or possibly additional quantum dots~\cite{Flensberg2011,Hyart2013,Aasen2016,Hell2016,Landau2016,Plugge2017,Karzig2017,Schrade2018}. In these schemes, the charging energy $E_C$ or a possible tunnel coupling between Majorana zero modes are used to lift the topological ground-state degeneracy in order to perform readout. In addition, by adiabatically changing the effective charging energies and/or tunnel couplings one can perform topologically protected qubit operations, equivalent to braiding in real space.

A different but non-protected approach uses tunnel coupling between two Majorana modes in a setup with two superconducting islands connected via a Josephson junction and placed within a resonant cavity \cite{Ginossar2014,Yavilberg2015}. It exploits the fact that the energy spectrum becomes insensitive to the local potential when the Josephson energy is sufficiently large, $E_J \gg E_C$ , eliminating dephasing due to local potential fluctuations, an idea borrowed from transmon qubits \cite{Koch2007}. The system considered in Refs.~\cite{Ginossar2014,Yavilberg2015}, termed the \emph{topological transmon}, has a spectrum which is approximately that of a doubly degenerate harmonic oscillator with frequency $\omega_p$ equal to the Josephson plasma frequency. Each single oscillator mode of the system contains a two-level system associated with the occupation of the subgap level formed from the Majorana edge modes  straddling the junction.

\begin{figure}
	\includegraphics[width = 0.5\textwidth]{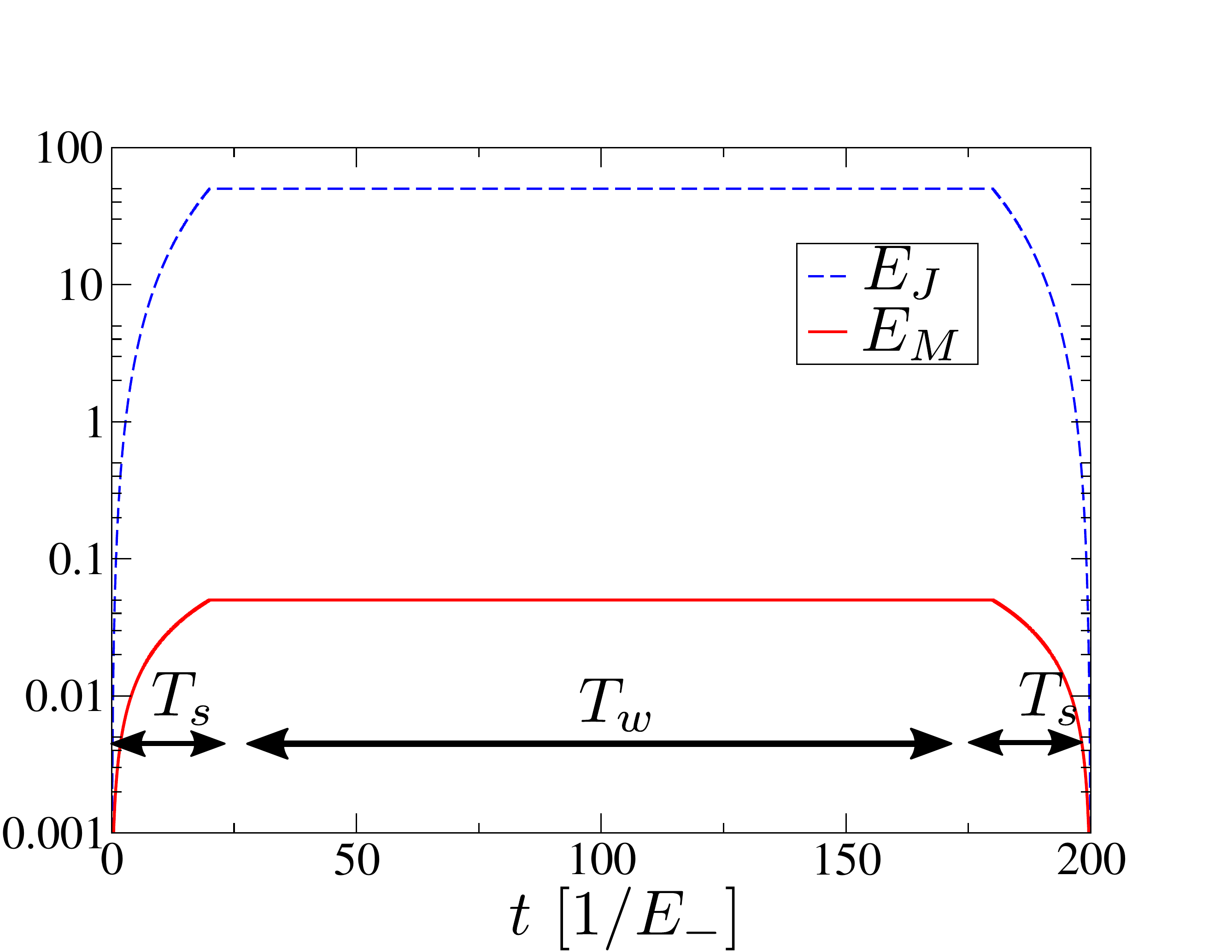}
	\caption{\label{fig:protocol}(\emph{Color online}) A protocol for qubit operation consisting of three stages of time-variation of the couplings $E_J(t)$ and $E_M(t)$. In the first stage, which occurs over a switching time $T_s$, the couplings are increased from zero, to values satisfying $E_J \gg E_- \gg E_M$. In the second stage, which occurs over a waiting time $T_w$, the couplings are held constant. In the final stage they are returned to zero.}
\end{figure}

In this paper, we discuss a device similar to the topological transmon, but instead of relying on readout in a basis of transmon states using a resonant circuit, we consider the possibility of initializing and reading out in the charge basis. Our system is shown in Fig.~\ref{fig:device}. It consists of two topological islands (blue) connected by a gate-controllable tunnel bridge (green), which gives rise to both the usual Josephson coupling $-E_J \cos \phi$ associated with tunneling of Cooper pairs across the junction as well as a coupling $i \gamma_2 \gamma_3 E_M \cos \frac{ \phi}{2}$ associated with tunneling of single electrons, mediated by the inner Majorana states. In addition, there are local gates controlling the potentials on the islands $V_L, V_R$. Another pair of gates is used to create trivial superconducting regions  (orange) which buffer the topological superconductors against the tunnel junction, so that the inner pair of Majorana modes $\gamma_2, \gamma_3$ remain spatially separated while maintaining a significant Josephson coupling $E_J$. In our proposed scheme, illustrated in Fig.~\ref{fig:protocol}, the islands are disconnected at the beginning and end of the protocols ($E_M = E_J = 0$), inducing Coulomb blockade on each island and making the charges of the left and right islands good quantum numbers. The qubit is initialized in a charge eigenstate $(\hat{N}_L - \hat{N}_R)|n \rangle = n |n\rangle$  (where $\hat{N}_L$ and $\hat{N}_R$ are charge operators for the left and right islands) with $n$ depending on the potentials $V_R, V_L$ (e.g. we may consider initialization in the state $|0 \rangle$). During the protocol, the tunnel coupling between islands is switched on via the middle gate, introducing Josephson couplings $E_J, E_M \neq 0$. The qubit is then operated as a topological transmon, which requires $E_J \gg E_- \gg E_M$ where $E_-$ is a charging energy associated with the difference between the number of electrons on the islands (so that the corresponding capacitive term in the Hamiltonian is $E_-(\hat{N}_L - \hat{N}_R)^2$). In this stage, the individual parities of the islands provide a spin-$\frac{1}{2}$ degree of freedom, within which the Majorana coupling $E_M$ generates coherent oscillations and thereby leads to a rotation of the qubit. At the end of the protocol, the tunnel junction is pinched off, restoring Coulomb blockade and the charge sensor is used to perform a measurement of the charge $\hat{N}_L$ on the left island.



A related scheme was suggested in Ref.~\onlinecite{Aasen2016} for validation of the topological qubit which was discussed in that work. Indeed the present proposal should be viewed in the same way. Inspired by the experimentally rather simple layout of the device, we wish to investigate how much can be learned about the coherence of the parity degree of freedom using techniques well-established in spin qubits. However, instead of spin-to-charge conversion \cite{Petta2005} \color{black}   we implement parity-to-charge conversion in order to read out the state of our system. We describe how to characterize the qubit properties, establish that there is a zero-energy state on each island and determine how the coupling between these can be controlled by the gates. However, it is not the aim of this work to suggest methods to experimentally prove that the zero-energy states are topological Majorana bound states. In fact, the experiments we propose can in principle also be carried out if the zero-energy states are Andreev bound states which remain fixed to near zero energy throughout the experiment, whether accidentally or by careful fine-tuning. We shall also discuss the potential obstacles to charge readout in our system, which we may consider to be indicative of similar schemes in future Majorana-based devices which may also need to be  operated without topological protection.\color{black} 

\begin{figure}
	\includegraphics[width = 0.5\textwidth]{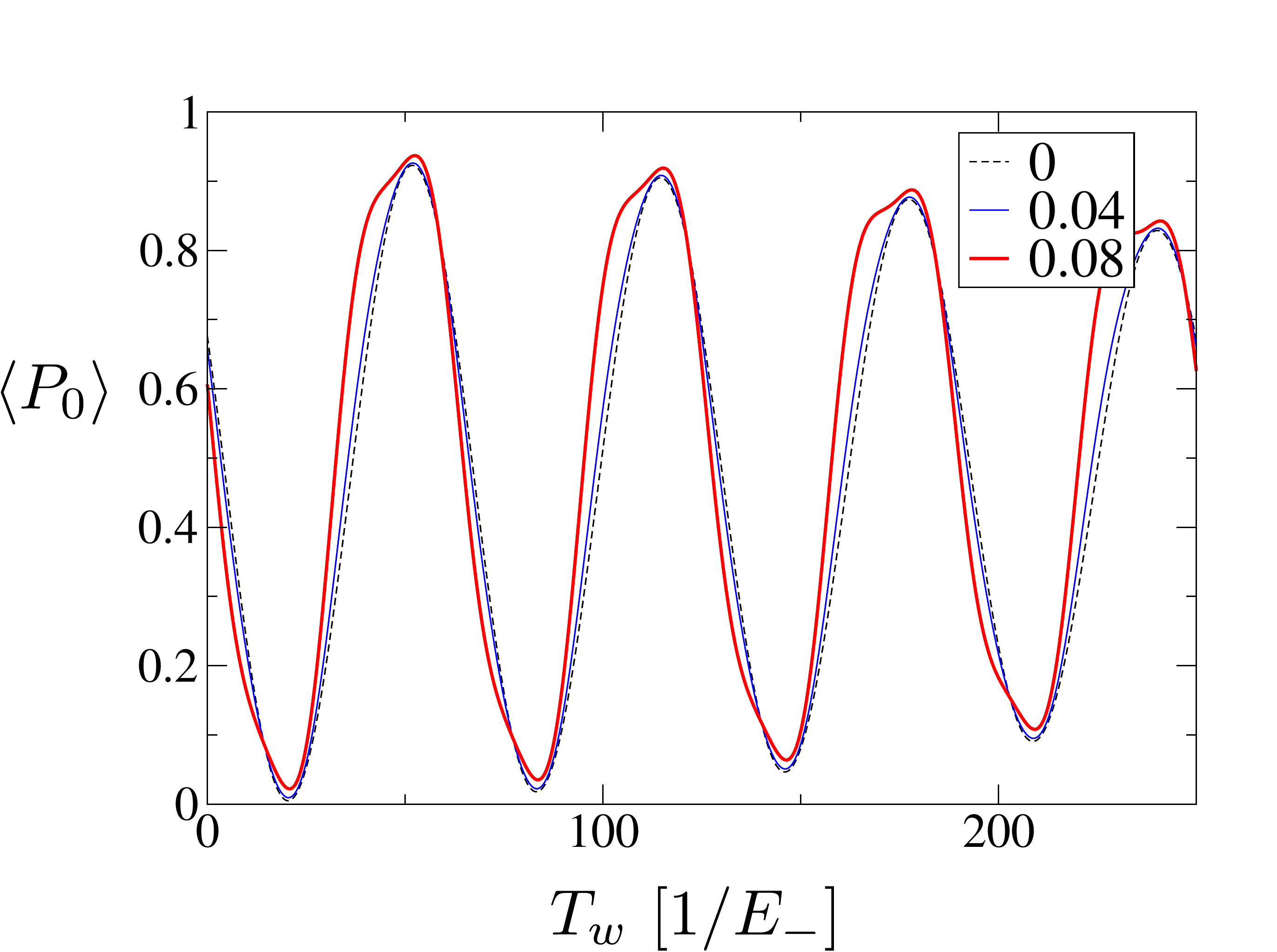}
	\caption{\label{fig:res}(\emph{Color online}) Our results for the coherent oscillations observed by charge sensing at the end of the protocol illustrated in Fig.~\ref{fig:protocol}, accounting for fluctuations $\delta E_M$ and $\delta V = \delta V_L - \delta V_R$. We calculate results for the particular case of a noise spectrum $\langle \langle \delta E_M(t) \delta E_M(0) \rangle \rangle = \langle \langle \delta E_M^2 \rangle \rangle = (0.02 E_M)^2$ with the average $E_M = 0.05 E_-$, and $\langle \langle \delta V(t) \delta V(0) \rangle \rangle = \langle \langle \delta V^2 \rangle \rangle =  0$ (black, dashed), $(0.04 E_-)^2$ (blue) and $(0.08 E_-)^2$ (red, bold).}
\end{figure}

In the remainder of this work we shall first describe the measurement of the ground state properties of the two-island system via charge sensing in the space of local gate voltages $(V_L, V_R)$ before discussing the dynamics of the multilevel system.
We present calculations of the visibility of coherent oscillations resulting from our proposed protocol accounting for fluctuations in the electrostatic environment beyond the Markov approximation and thus account for a general frequency-dependent noise spectrum. We describe several important sources of decoherence, and conclude that while fluctuations in $E_M$ resulting from noise in the middle gate lead to pure dephasing at all times when $E_M$ is nonzero, the influence of fluctuations in the potentials $\delta V(t) = \delta V_L(t) - \delta V_R(t)$ is considerably more subtle and depends crucially on the noise spectrum $\mathcal{S}(\omega) = \int{ \langle\langle \delta V(t) \delta V(0) \rangle\rangle e^{i \omega t} dt}$. When the system is in the transmon regime, the main contribution to decoherence consists of fluctuations with $\omega = \omega_p$ which lead to linear decay of the oscillations for short times. However, accounting for the experimental limits on the time $T_s \gtrsim 1\text{ ns}$ required for the switching on of the tunnel barrier, we find that low-frequency noise may lead to significant decoherence during (and only during) the conversion stages at the beginning and end of the protocol in a manner which is extremely sensitive to the dynamical evolution of the many-body wavefunction. The mechanism we describe does not contribute to the decay of the oscillations but rather introduces corrections to the expected value $\langle P_0 \rangle$ of the projection operator onto a charge state $P = |0 \rangle \langle 0 |$ which are periodic in the waiting time $T_w$. We illustrate these results in Fig.~\ref{fig:res},  where we have accounted for both $\delta V(t)$ and $\delta E_M(t)$ fluctuations in the quasi-static regime, assuming that the correlation time is long compared to the period of coherent oscillations but short on the time-scale of a complete experiment involving many individual measurements, so the conditions $\langle \langle \delta V(t) \delta V(0) \rangle \rangle = \langle \langle \delta V^2 \rangle \rangle$, $\langle \langle \delta E_M(t) \delta E_M(0) \rangle \rangle = \langle \langle \delta E_M^2 \rangle \rangle$ are fulfilled for each individual measurement and the average is taken over random variations in the static parameters. Fluctuations in $E_M$, which are associated with pure dephasing, lead to a decay of the oscillations $\delta \langle P_0 (T_f) \rangle = -4 \langle \langle \delta E_M^2 \rangle \rangle T_w^2$ while fluctuations in $V(t)$ result in the distortion of the oscillations. We show results for $\sqrt{ \langle \langle \frac{\delta E_M^2}{E_M^2} \rangle \rangle} = 0.02$, and various values of $\sqrt{ \langle \langle \frac{\delta V^2}{E_-^2} \rangle \rangle} = 0, 0.04, 0.08$. It should be noted that the value of the oscillations is smaller than  unity at $T_w = 0$ due to the finite switching time $T_s = 20/E_-$.
\color{black}

The remainder of the paper is structured as follows: in Section II we introduce the model describing our system, and describe its level spectrum and some properties which can be probed via charge sensing. In Section III we consider protocols for pulsing the gates to generate coherent oscillations. In Section IV we calculate the corrections due to the visibility arising from decoherence due to coupling to classical fluctuations of the local potential. Finally, in Section V we present a summary of our results and conclude.

\section{Spectral properties of the double island qubit}

We model our system (shown in Fig.~\ref{fig:device}) by a Hamiltonian consisting of single-electron number operators $\hat{N}_L, \hat{N}_R$ and the superconducting phase difference $\phi = \phi_L - \phi_R$:
\begin{gather}
H = \frac{1}{2} \sum_{i,j = L,R}{ C^{-1}_{ij} \hat{N}_i \hat{N}_j } - \sum_{i = L, R}{ V_i \hat{N}_i} \nonumber \\
- E_J \cos \phi  - i E_M \gamma_2 \gamma_3 \cos  \frac{\phi}{2}  \ \ .
\end{gather}

Since the two-island system is chemically isolated from the environment, the total charge $\hat{N}_L + \hat{N}_R = N$ is conserved. It is convenient to express the Hamiltonian in terms of the relative number operator $\hat{n} = \hat{N}_L - \hat{N}_R = 4i \frac{\partial}{\partial \phi}$, so the Hamiltonian reads
\begin{gather}
H = E_+ ( N - N_g)^2 + E_- ( \hat{n} - n_g)^2 \nonumber \\
- E_J \cos \phi + i E_M \gamma_2 \gamma_3 \cos \frac{ \phi}{2} \ \ ,
\label{Hamil2}
\end{gather}
where we have introduced the total charging energy
$E_+ = \frac{ C^{-1}_{LL} + C^{-1}_{RR} + 2C^{-1}_{LR}}{8}$, the relative charging energy  $E_- = \frac{C^{-1}_{LL} + C^{-1}_{RR} - 2C^{-1}_{LR}}{8} $ and the dimensionless gate voltages $N_g = \frac{ V_R + V_L}{4 E_+}\ \ , \ \ n_g = \frac{ V_L - V_R}{4 E_-} + \frac{C^{-1}_{LL} - C^{-1}_{RR}}{2 ( C^{-1}_{LL} + C^{-1}_{RR} - 2 C^{-1}_{LR})} N$. For simplicity, in future we will consider the case when $C^{-1}_{LL} = C^{-1}_{RR}$.
\color{black}
Due to the existence of zero-energy modes, the total number $N$ may be either even or odd in the absence of excitations of quasiparticles above the superconducting gap. \color{black} We also assume that the size of each island is large compared to the spatial extent of the Majorana bound states, which implies that the outer Majorana modes are not coupled to the rest of the system.

\begin{figure}[t]
	\includegraphics[width = 0.55\textwidth]{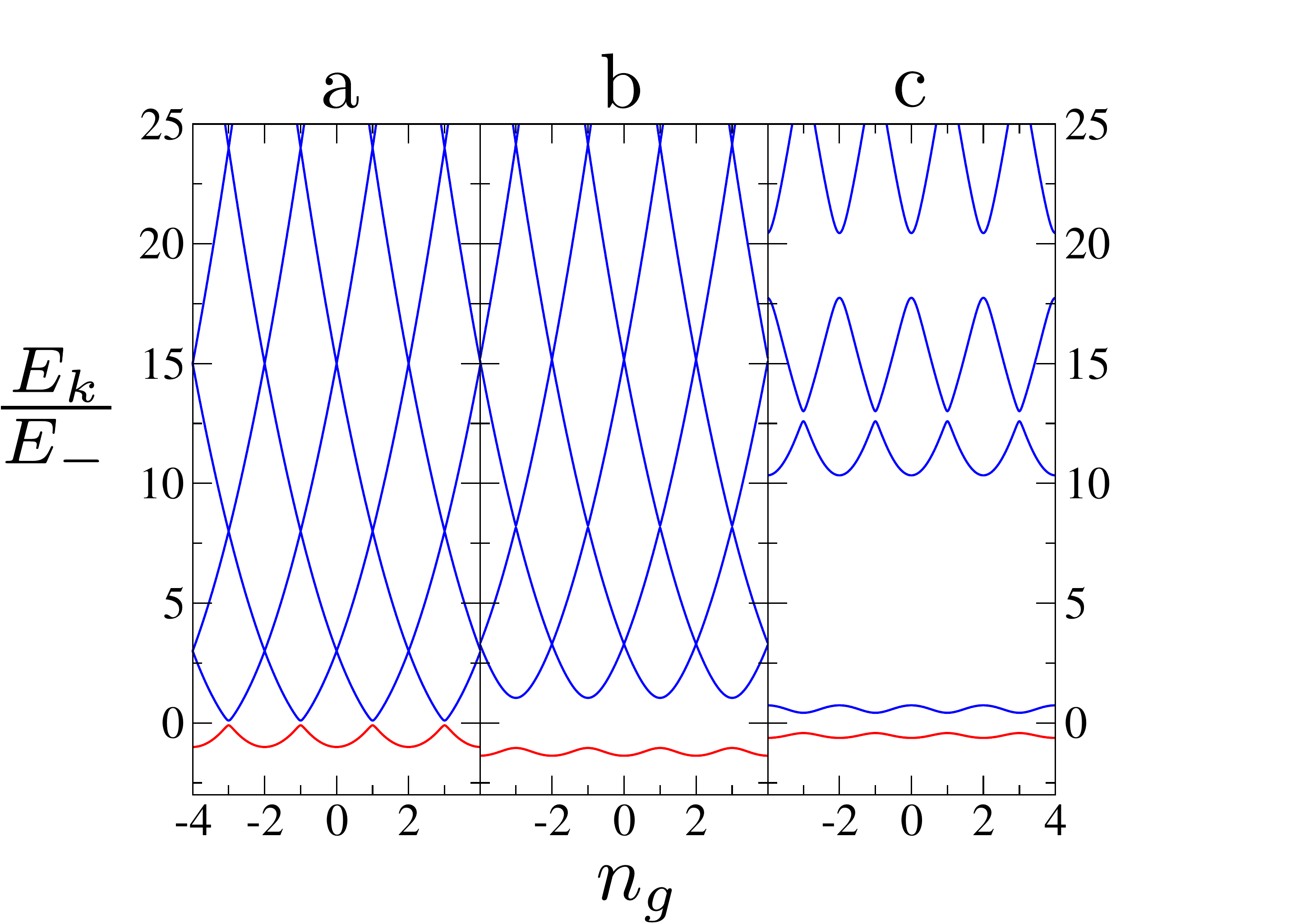}
	\caption{\label{fig:paras}( The dispersion of energy levels $E_k$ as a function of $n_g$ for the situations: a) $E_J = 0$, $\frac{E_M}{E_-} = 0.2$, b) $\frac{E_J}{E_-} = 0.5$, $\frac{E_M}{E_-} = 2$, and c) $\frac{E_J}{E_-} = 10$, $\frac{E_M}{E_-} = 0.5$. \emph{Color online}: The energy of the ground state $E_0$ is highlighted in red.}
\end{figure}

We plot the spectrum $E_k(n_g)$ of the two-island system in Fig.~\ref{fig:paras} for three cases:  (1)  $E_J, E_M \ll E_-$, (2)  $E_J \ll E_M \sim E_-$ and (3) $E_M \ll E_- \ll E_J$.
We assume $N$ is even and treat the first term in (\ref{Hamil2}) involving the total capacitance as a constant offset and subtract it in all cases. In case (1), which is plotted in Fig.~\ref{fig:paras}a for $E_J = 0$, $E_M = 0.2 E_-$, neighbouring parabolic bands correspond to states of definite charge in which a single electron is transferred across the junction, and the charge difference $n = N_L - N_R$ in the ground state changes by two at the crossings of the lowest parabolas. In cases (2) and (3), plotted in Figs.~\ref{fig:paras}b and \ref{fig:paras}c respectively, we may observe that the dispersion of the lowest band is suppressed. In case (2), plotted for $E_J = 0.5 E_-$, $E_M = 2E_-$ the major anticrossings occur at odd values of $n_g$ and between charge states with odd and even parity; the lowest band is gradually separated from the excited states with increasing $E_M$. In case 3, plotted for $E_J = 10E_-$, $E_M = 0.5 E_-$, a pair of closely spaced, weakly dispersing bands emerges at the bottom of the spectrum.

\begin{figure}
	\includegraphics[width = 0.47\textwidth]{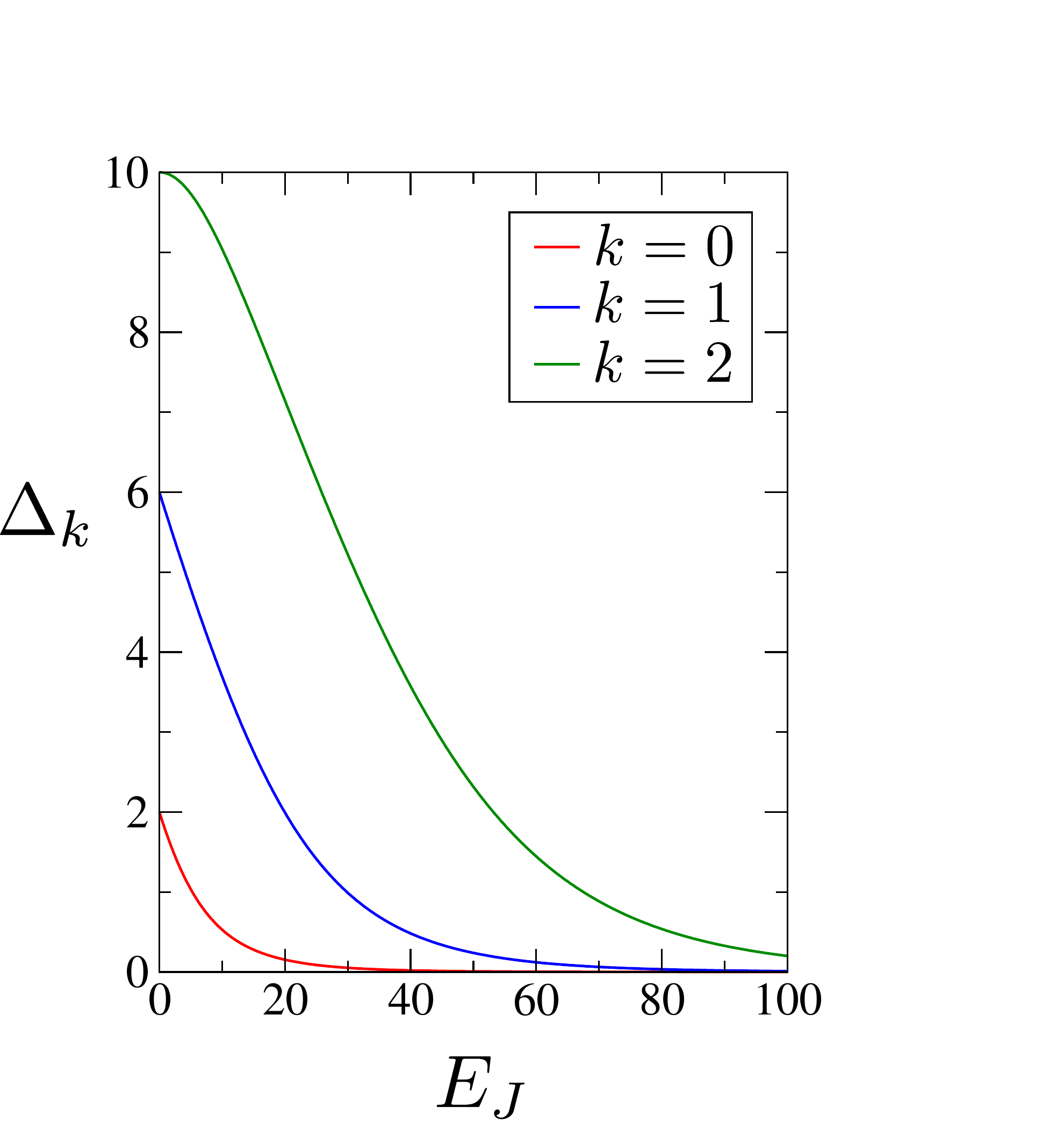}
	\caption{\label{fig:amp}(\emph{Color online}) The amplitude of the oscillations in the energy of neighbouring bands $\Delta_k = \frac{ E_{2k+1}(n_g) - E_{2k}(n_g)}{2}$ at their maximum value ($n_g = 0$) for (from bottom to top) $k = 0$ (red), $k = 1$ (blue) and $k = 2$ (green), for $E_M = 0$. Units of energy are chosen so that $E_- = 1$.}
\end{figure}

\begin{figure}
	\includegraphics[width = 0.5\textwidth]{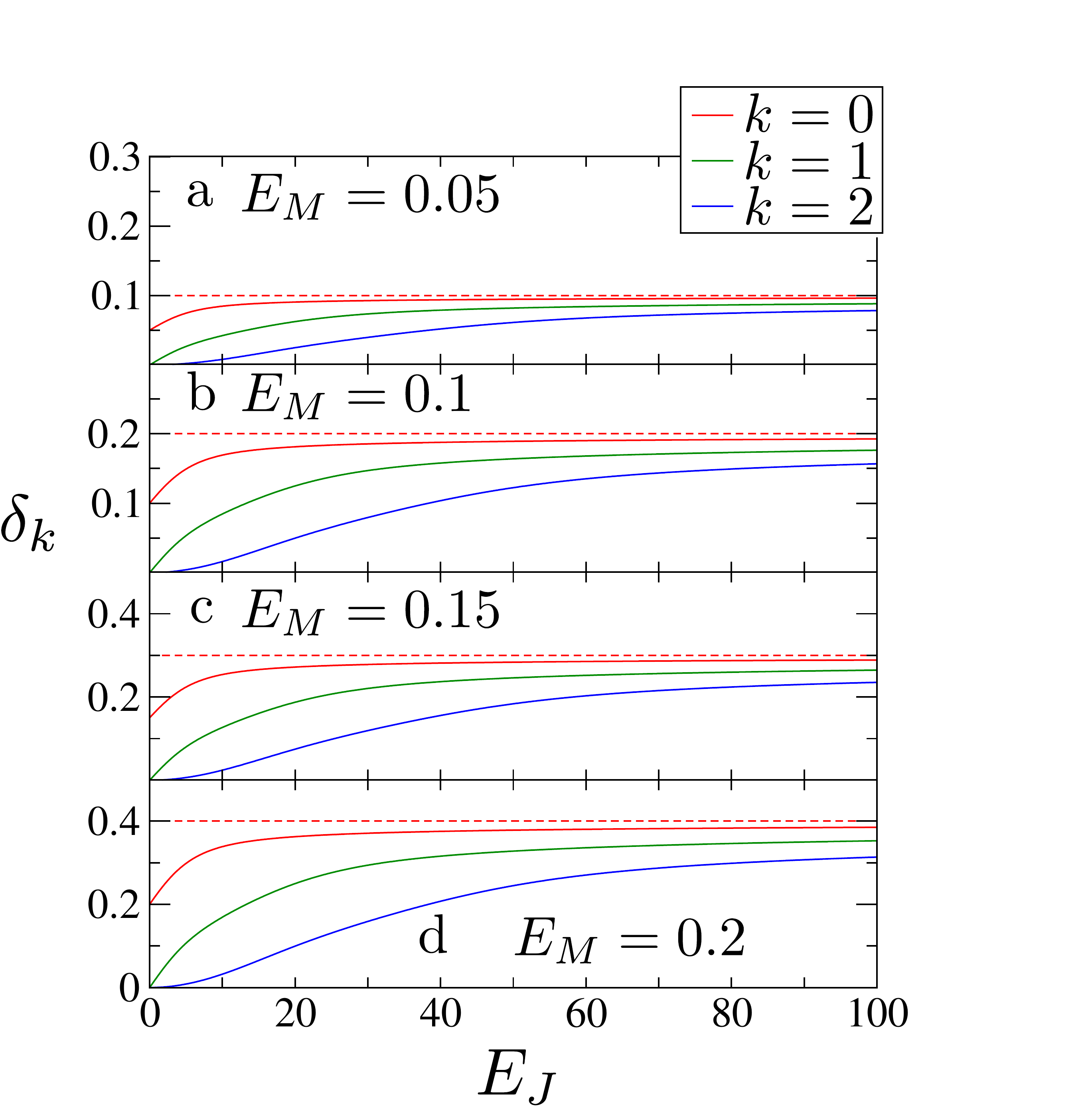}
	\caption{\label{fig:msplitting}(\emph{Color online}) The splitting between pairs of bands $ \delta_k = E_{2k+1}(n_g)-E_{2k}(n_g)$ at their minimum values (corresponding to $n_g = 1$), shown in the solid lines for (top to bottom) $k = 0$ (red), $k = 1$ (green), $k = 2$ (blue). Units of energy are chosen so that $E_- = 1$. The dashed red line indicates the energy $2E_M$.}
\end{figure}

Since the total number operator $\hat{N}_L + \hat{N}_R$ commutes with the Hamiltonian, the many-body wavefunction is simply a function of the phase difference and satisfies the periodic (antiperiodic) boundary conditions $\psi(\phi + 4\pi) = (-1)^N \psi(\phi)$ for even (odd) $N$. As a result, the spectrum is periodic, $E_k(n_g + 2) = E_k(n_g)$. In the regime $E_J \gg E_- \gg E_M$, $\psi(\phi)$ is localized around the minima of the potential energy $-E_J \cos \phi$, and a standard approach \cite{Ginossar2014} involves replacing it with a harmonic oscillator potential around one of the minima $\phi \approx 0$. After a gauge transformation $\psi \rightarrow e^{-\frac{ i n_g \phi}{4}}\psi$  the Hamiltonian (\ref{Hamil2}) (omitting the constant term containing $N$) takes the harmonic oscillator form
\begin{gather}
H = -16 E_- \frac{ \partial^2}{\partial \phi^2} + \frac{E_J}{2} \phi^2
\end{gather}
with the dependence on $n_g$ having been absorbed into new boundary conditions. Since the wavefunction is localized around $\phi \approx 0$, it becomes insensitive to the boundary conditions and the spectrum consists of a series of harmonic oscillator levels with level spacing equal to the frequency $\omega_p = \sqrt{32 E_J E_-}$. 

The lowest states in the spectrum correspond to quantized oscillations of the superconducting phase difference $\phi$ at the plasma frequency $\omega_p$. Since the depth of the potential energy is $2 E_J$, the number of oscillator levels may be approximated by
\begin{gather}
N_{osc} = \sqrt{ \frac{E_J}{8 E_-}} \ \ ,
\end{gather}
while the remaining excited states are approximately charge eigenstates which are  perturbed by the Josephson energy. We note that our expression depends on the mutual capacitance $(C^{-1}_{LR})$ of the two islands which reduces $E_- = \frac{C^{-1}_{LL} - C^{-1}_{LR}}{4}$. Since charge dispersion is suppressed either for $E_J \gg E_-$ or $E_M \gg E_-$, it is clear that, aside from the mutual capacitance often being a significant effect in experiment, it may be crucial in driving experiments into the transmon regime since $E_-$ may be considerably reduced by the capacitive coupling between the islands when it is close to the self-capacitance of the individual islands.

In the case $E_M = 0$ where parity is a good quantum number, the spacing between opposite parity states within successive pairs of bands collapses as the system is driven into the transmon regime, with the higher pairs of levels becoming degenerate at successively larger values of $E_J$. We have plotted the maximum splitting $\Delta_k = \frac{E_{2k+1}(n_g = 0) - E_{2k}(n_g = 0)}{2}$  for $E_M = 0$ in Fig.~\ref{fig:amp}, for the bands corresponding to $k = 0$ (red), $k = 1$ (blue) and $k = 2$ (green) as a function of $E_J$. 

As $E_M$ is switched on, the oscillator levels $E_{2k+1}$ and $E_{2k}$, initially of definite and opposite parity, become mixed by the Majorana coupling $i E_M \gamma_2 \gamma_3  \cos \frac{ \phi}{2}$ in Eq. (\ref{Hamil2}), which results in an energy splitting within each pair which is dependent on $E_M$.  In Fig.~\ref{fig:msplitting} we plot the minimum splitting $\delta_k = E_{2k+1}(n_g = 1) - E_{2k}(n_g = 1)$ after the introduction of the Majorana coupling within each doublet as a function of $E_J$ for various values $\frac{E_M}{E_-} = 0.05$ (a), 0.1 (b), 0.15 (c), and 0.2 (d), and $k = 0$ (red), $k = 1$ (green) and $k = 2$ (blue). The dashed red line indicates an energy of $2E_M$. For all levels, the minimum splitting converges to $2 E_M$ as the system is driven into the transmon regime.

\begin{figure}
	\begin{tabular}{c}
\includegraphics[width =0.5\textwidth]{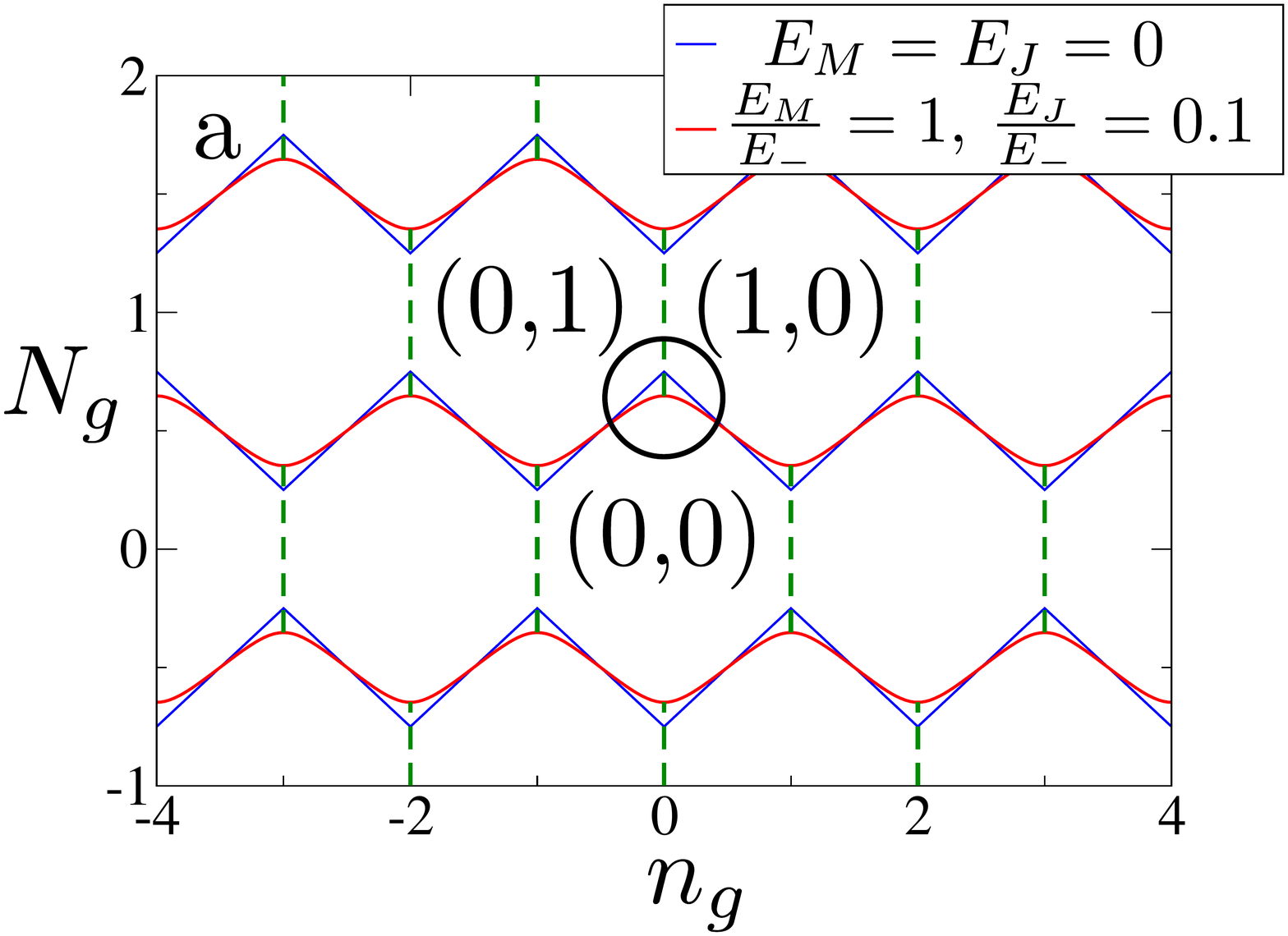}
\\
\includegraphics[width = 0.5\textwidth]{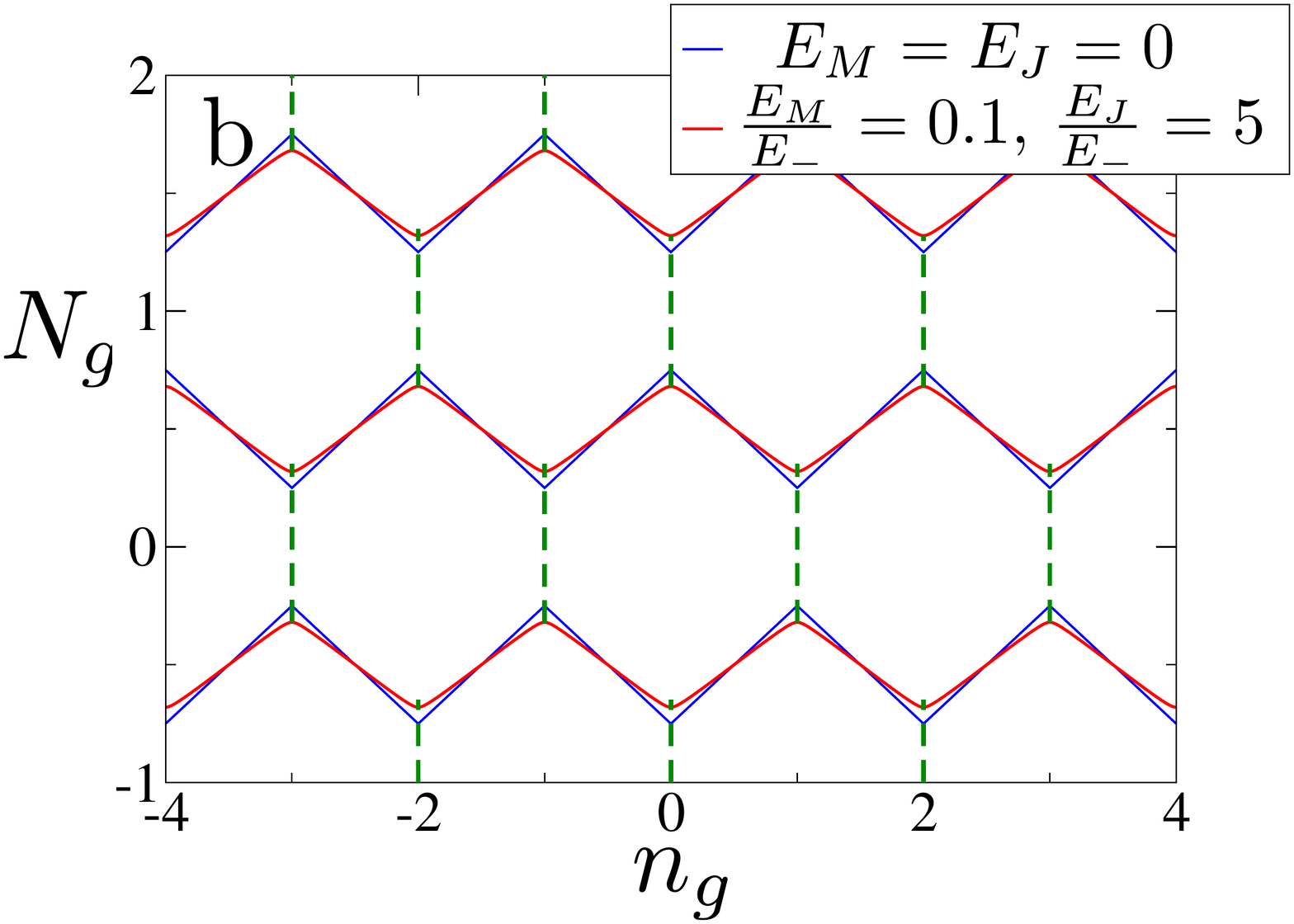}
\end{tabular}
\caption{\label{fig:honey}(\emph{Color online}) The charge stability diagram of the two-island system, which consists of regions separated by phase boundaries at which the average charge $(N_L, N_R)$ on either of the two islands exhibits a step. The phase boundaries separating states of different total number $N$ are shown in the curved red lines for
for parameters (a)  $\frac{E_M}{E_-} = 1$, $\frac{ E_J}{E_-} = 0.1$ and  (b) $\frac{ E_M}{E_-} = 0.1$, $\frac{E_J}{E_-} = 5$, and in the blue straight lines for $E_J = E_M = 0$. The dashed green lines indicate the boundary between phases at which the charge is increased by a single electron on one island and decreased by a single electron on the other. The circle in panel b indicates the region in which the phase boundary may be described by Eq. (\ref{phaseboundary2}).
}
\end{figure}

In our proposed geometry (Fig.~\ref{fig:device}), measurements of the qubit are performed by charge sensing rather than transport through the islands.  When the parameters $(n_g, N_g)$ are varied, the measurement of the average charge on the left island becomes a probe of the charge structure of the ground state, which is controlled by the ratios of couplings $E_J/E_-$ and $E_M/E_-$. It is therefore important to determine the circumstances under which these couplings may be extracted via charge sensing.

We shall briefly consider an extension of our setup which allows the charge states to be measured for different values of $N_g$ and $n_g$ due to a coupling between one or both islands and an electron reservoir. We assume that this tunnel coupling is weak enough that we can ignore the associated broadening of the charge states. In addition, we assume that temperature is low enough and that the sweeping rates of $N_g$ and $n_g$ are slow enough, such that the system remains in the ground state. The charge stability diagram is plotted in Fig.~\ref{fig:honey} for particular values of $E_J$ and $E_M$ with fixed $E_+$ and $E_-$. The blue lines indicate the situation when $E_J = E_M = 0$, while red lines correspond to $E_J, E_M \neq 0$. In the former situation, the charge stability diagram consists of hexagonal regions with areas given by $A = 2C^{-1}_{LL}/(C^{-1}_{LL} + 2 C^{-1}_{LR})$ and therefore fixed simply by the capacitances. Sharp vertical phase boundaries appear at the transitions in which a single electron is transferred between the islands while zigzag boundaries separate horizontal regions in which a single electron is added or removed from the entire system. The areas of the distinct phases therefore reflects the presence of a zero energy subgap state which allows for a change in the fermionic parity of either a single island $(-1)^{\hat{N}_L} = -i \gamma_1 \gamma_2$, $(-1)^{\hat{N}_R} = -i \gamma_3 \gamma_4$ or of the whole system, $(-1)^{\hat{N}_L + \hat{N}_R} = -\gamma_1 \gamma_2 \gamma_3 \gamma_4$. \color{black}

With a nonzero value of either $E_M$ or $E_J$, the boundaries separating horizontal regions with different $N$ remain sharp but become rounded in both cases. When $E_M$ is nonzero, the inner Majorana modes $\gamma_3, \gamma_4$ are coupled and the individual parities $i \gamma_1 \gamma_2$, $i \gamma_3 \gamma_4$ no longer commute with the Hamiltonian, leading to the disappearance of the vertical boundaries seen for $E_M = 0$. For the case $E_J \ll E_-, \frac{E_-^2}{E_M}$ the height of the phase boundary between $N = 0$ and $N = 1$ states near $n_g = 0$ (indicated by the dashed line above the circle in Fig. \ref{fig:honey}a)  is given by
\begin{gather}
N_g = \frac{1}{2} +\frac{E_-}{2 E_+} - \sqrt{ ( \frac{E_- n_g}{ E_+})^2 + (\frac{ E_M}{2 E_+})^2 } \ \ ,
\label{phaseboundary2}
\end{gather}
and reflects the anticrossings at the odd integer values of $n_g$ observed in Fig.~\ref{fig:paras}a. As $E_M$ is reduced to zero, the phase boundary develops a cusp at $n_g = 0$. This cusp remains even when the condition $E_J \ll E_-$ is not fulfilled, but the height of the phase boundary is reduced quadratically in $E_J$. This situation is illustrated in Fig. \ref{fig:honey}b, with parameters $E_M = 0.1E_-$ and $E_J = 5 E_-$.

\begin{figure*}[t!]
\includegraphics[width=\textwidth]{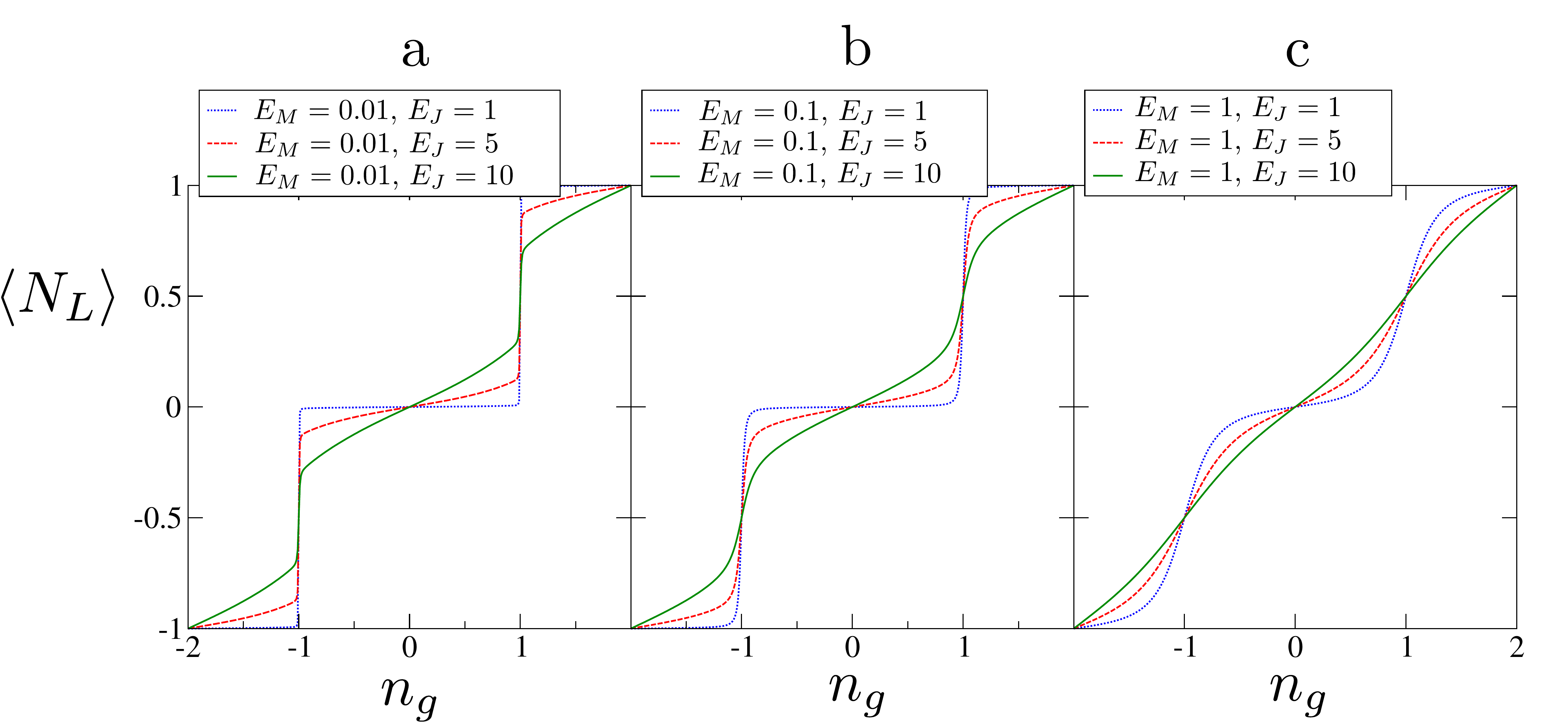}
\caption{\label{fig:steps}(\emph{Color online}) The average charge on one island $\langle N_L\rangle$ as a function of $n_g$ for (a) $E_M = 0.01$, (b) $E_M = 0.1$ and (c) $E_M = 1$ in units of $E_-$, for various values of $E_J$.}
\end{figure*}
\twocolumngrid

From Eq.~\eqref{Hamil2}, we may see that the average charge on the left island, $\langle \hat{N}_L \rangle = \frac{1}{2} ( \langle \hat{n} \rangle + N)$, may be related to the ground state energy $E_G$ via $\frac{dE_G}{dn_g} = 2 E_- ( \langle \hat{n} \rangle - n_g)$. Thus, within a region of fixed total number $N$, $\langle N_L\rangle$ will be constant as a function of $N_g$ for fixed $n_g$. The charge on the left island is plotted as a function of $n_g$ for several values of $E_J/E_-$, $E_M/E_-$ in Fig.~\ref{fig:steps}. In  situations illustrated in (a) and (b), the charge exhibits well-defined steps at odd integer values of $n_g$ which are associated with the extrema of the lowest band of the spectra illustrated in Fig.~\ref{fig:paras}. When the  islands are individually Coulomb blockaded, anticrossings between states of opposite parity occur at odd integer values of $n_g$, and at these values the separation in energy between the lowest two states of the same parity is equal to $8 E_-$. Thus charge fluctuations associated with $E_J$ may be treated perturbatively when $E_J \ll 8 E_-$. If, additionally, $E_M = 0$, the parity of each island is fixed and the average charge $\langle \hat{N}_L\rangle$ exhibits a vertical jump as the ground state changes abruptly at the level crossing between states of opposite parity. For nonzero $E_M$, this jump remains visible as long as $E_M \ll E_-$. To lowest order in $\frac{E_J}{E_-}$ the average charge for $n_g \approx 0$ is given by
\begin{gather}
\langle  \hat{N}_L \rangle =
\left( \frac{E_J^2}{256 E_-^2} + \frac{E_M^2}{16 E_-^2}\right) n_g
\end{gather}
while for $n_g \approx 1$ we have
\begin{gather}
\langle \hat{N}_L \rangle = \frac{1}{2} + \frac{E_-(n_g - 1)}{ \sqrt{( \frac{E_M}{2})^2 + 4 E_-^2 ( n_g - 1)^2 } } \ \ ,
\end{gather}
and the slope of the step at $n_g = 1$ is equal to $\frac{2 E_-}{E_M}$.

In the topological transmon regime,
$E_J \gg E_- \gg E_M$, the many-body wavefunction may be approximated by harmonic oscillator states in the phase representation and the spectrum may be solved by standard methods \cite{Ginossar2014}. We find that
\begin{gather}
\langle \hat{N}_L \rangle = \frac{n_g}{2} + \frac{ \pi t^2 \sin \pi (n_g-1)}{ 32 \sqrt{ (E_M/E_-)^2 + t^2 \sin^2 \frac{ \pi (n_g-1)}{2}} }
\label{transmonformula}
\end{gather}
where
\begin{gather}
t = 64 \sqrt{ \frac{2}{\pi}} \left( \frac{E_J}{8E_-}\right)^{\frac{3}{4}} e^{- \sqrt{ 2E_J/E_-}} \ \ .
\end{gather}
Thus, both in the cases when  $E_J, E_M \ll E_-$, and $E_J \gg E_- \gg E_- t \gg  E_M$, sharp steps are visible at values of $n_g$ corresponding to anticrossings between different parity states. In the latter case, the large value of $E_J$ results in slopes $\frac{d \langle \hat{N}_L \rangle}{d n_g} > 0$ for even integer values of $n_g$ (as seen in Fig. \ref{fig:steps}a,b), while a vertical step survives at odd integer values of $n_g$, reflecting the fact that the term $-E_J\cos \phi$ in the Hamiltonian (\ref{Hamil2}) can only generate fluctuations in the number of Cooper pairs on each island without affecting the fermionic parity.

From Eq. (\ref{transmonformula}) we observe that $\frac{d \langle \hat{N}_L \rangle}{d n_g}$ is maximum at odd values of $n_g$ and minimum at even values. For $n_g$ odd the slope is given by
\begin{gather}
\frac{ d \langle \hat{N}_L \rangle}{dn_g} = \frac{1}{2} + \frac{ \pi^2 t^2 E_-}{32 E_M}
\end{gather}
while for $n_g$ even
\begin{gather}
\frac{ d \langle \hat{N}_L \rangle}{dn_g} = \frac{1}{2} - \frac{ \pi^2 t^2}{ 32 \sqrt{( E_M/E_-)^2 + t^2}} \ \ .
\end{gather}
Comparison of the slopes at odd and even integer values of $n_g$ therefore allows the direct extraction of two distinct parameters $E_M$, $E_J$, indicating the existence of two independent couplings proportional to $\propto \cos \frac{\phi}{2}$ and $\propto \cos \phi$.

\section{Coherent oscillations}

\begin{figure}[t!]
\includegraphics[width = 0.5\textwidth]{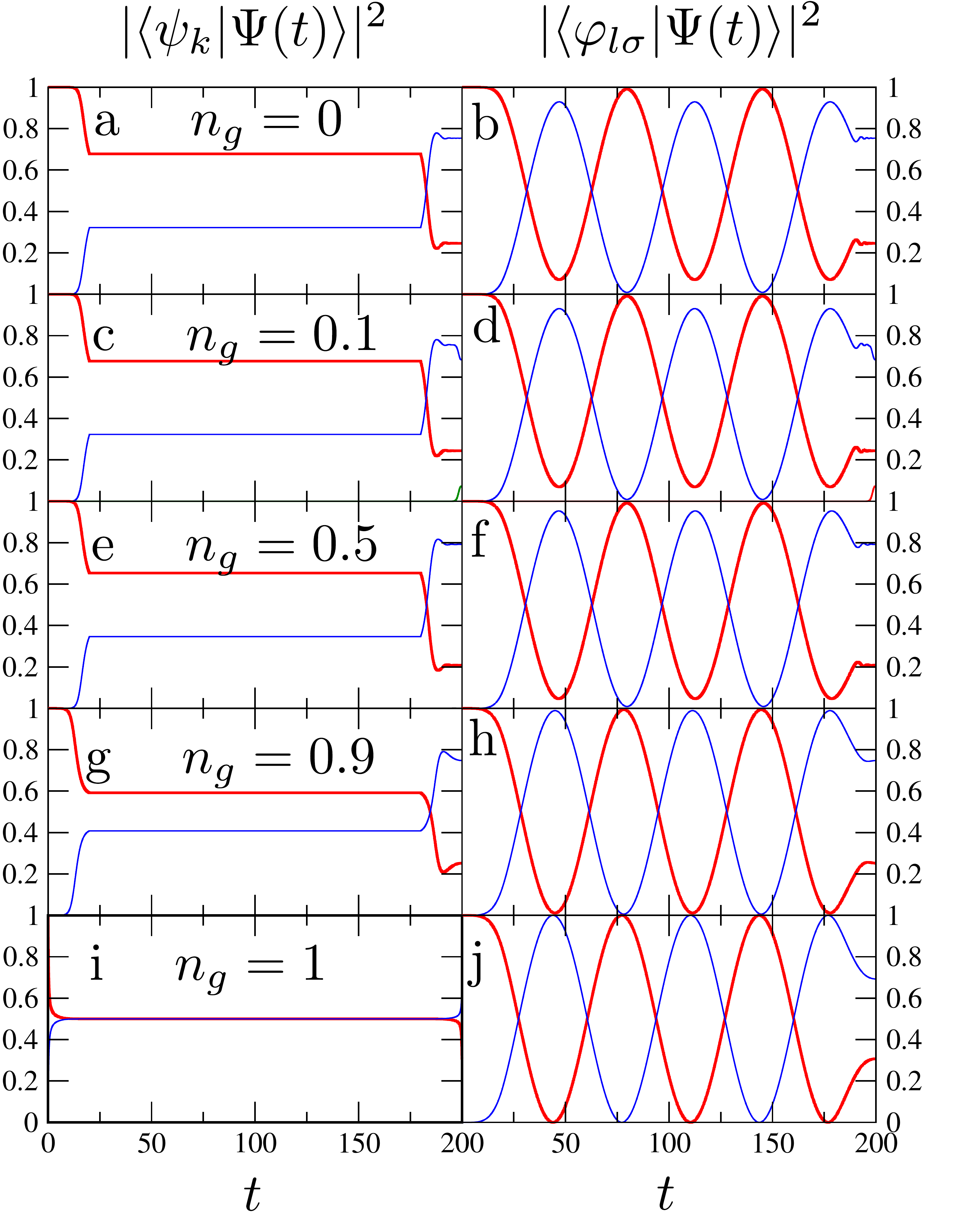}
\caption{\label{fig:osc}(\emph{Color online}) Coherent oscillations for various choices of $n_g = 0$ (a,b), $n_g = 0.1$ (c, d), $n_g = 0.5$ (e,f), $n_g = 0.9$ (g, h), $n_g = 1$ (i, j). The left panels (a,c,e,g,i) show the time evolution of the components of the wavefunction $\Psi(t)$ projected onto the instantaneous adiabatic eigenbasis of the \emph{full} Hamiltonian $H(t)$ (\ref{H0}), while the right panels (b,d,f,h,j) show the components in the instantaneous eigenbasis of the partial Hamiltonian $H_0(t)$ which excludes the Majorana coupling $\propto E_M$. The values of the inter-island couplings are taken to be $E_J = 50E_-$, $E_M = 0.05E_-$ during the waiting period, and the switching time is $T_s = \frac{20}{E_-}$.}
\end{figure}

As we saw in Fig.~\ref{fig:paras}c, the lowest states of the charge spectrum in the regime $E_M \ll E_- \ll E_J$ (i.e. the topological transmon) consists of a pair of harmonic oscillator levels which disperse weakly with $n_g$ and are separated from the excited states by the plasma freqency $\omega_p = \sqrt{32 E_J E_-}$. If the Majorana coupling $E_M$ is varied with time, coherent operations may be performed within the doublet of levels within the lowest mode of the transmon without exciting plasma oscillations. In this section, we will analyse the operation of a pulsing protocol in which both $E_J$ and $E_M$ are varied in time via pulsing of the three gates shown in Fig.~\ref{fig:device}, with initialization and readout occurring under Coulomb blockade $E_J, E_M = 0$. The protocol, which realizes a Ramsey interferometry experiment, is illustrated in Fig.~\ref{fig:protocol}.  The tunnel junction is initially closed, with thermal relaxation initializing the qubit in the lowest charge state, which we may assume to be $n = 0$ (and therefore $-1 < n_g < 1$). The tunnel junction is then opened over a switching period $0 < t < T_s$, driving the system into the topological transmon regime $E_J \gg E_- \gg E_M$. During this period, the instantaneous eigenstates of the time-dependent Hamiltonian $H(t)$ evolve continuously from states of definite charge into a spectrum of topological transmon states which consists of doublets of oscillator levels. During this period, the many-body wavefunction must be converted into an equal superposition of states within the lowest doublet. $E_J$ and $E_M$ are then fixed over a waiting period $T_s \leq t \leq  T_s + T_w$, and the quantum state undergoes coherent oscillations. In the final stage of the process, $T_s + T_w < t \leq T_f$, $E_J$ and $E_M$ are returned to zero, converting the quantum state into a superposition of the lowest two charge states. The charge sensor then performs a projective measurement of $\hat{N}_L$. In an optimal experimental situation, it is sensitive to only few $\hat{N}_L$ eigenstates.

In the remainder of the paper, we consider the expectation value of an outcome of the measurement which is equal to projection onto a given charge state, namely $|n = 0 \rangle$,
\begin{gather}
\langle P_0\rangle_{\text{final}}=\text{Tr} P_0 \rho(T_f)
\label{measurement}
\end{gather}
where $P_0 = |0 \rangle \langle 0|$ is the projection operator and $\rho(t)$ is the density matrix, which exhibits coherent oscillations as a function of the waiting time $T_w$. The visibility of oscillations is maximal under the conditions that 1) the many-body wavefunction remain within the subspace spanned by the lowest two instantaneous eigenstates of $H(t)$ during the process, and 2) the initial stage $0 < t < T_s$ results in its conversion into an equal superposition of the lowest two instantaneous eigenstates. If $E_J$ and $E_M$ were switched on too slowly, the quantum state would remain within the instantaneous ground state and avoid coherent oscillations. At the same time, the switching time $T_s$ must be sufficiently long as to avoid leakage to the higher excited states. At the start of the switching interval, the lowest two charge states are separated from the higher levels by a splitting of the order of $16 E_-$, and this splitting evolves into multiples of the plasma frequency $l\omega_p = l\sqrt{32 E_J E_-}$ at the end of the interval where $l$ is the oscillator index. At the same time, the matrix element of the Josephson coupling $-E_J(t) \cos \phi$  connecting the lowest doublet to the excited states varies between $-\frac{E_J}{2}$ and $-\frac{\omega_p}{\sqrt{2}}$. At the end of the switching interval the interaction $-E_J \cos \phi \sim \frac{E_J\phi^2}{2}$ only connects the lowest doublet to the states with oscillator index $l = 2$. It follows that the time $T_s$ must satisfy $T_s \gg \frac{E_J}{128 E_-^2} \gg \frac{ 1}{4\sqrt{2} \omega_p}$ (where the second inequality is implied by the transmon criterion $E_J \gg E_- \gg E_M$).  \color{black} Taking parameters $E_- = 10\mu\text{eV}$ and $E_J = 40 E_-$, this requires $T_s \gg 0.02 \text{ ns}$. In most experiments switching times are limited to $T_s > 1\text{ ns}$, so we expect that condition 1) will always be fulfilled.

\begin{figure}[t!]
\includegraphics[width = 0.48 \textwidth]{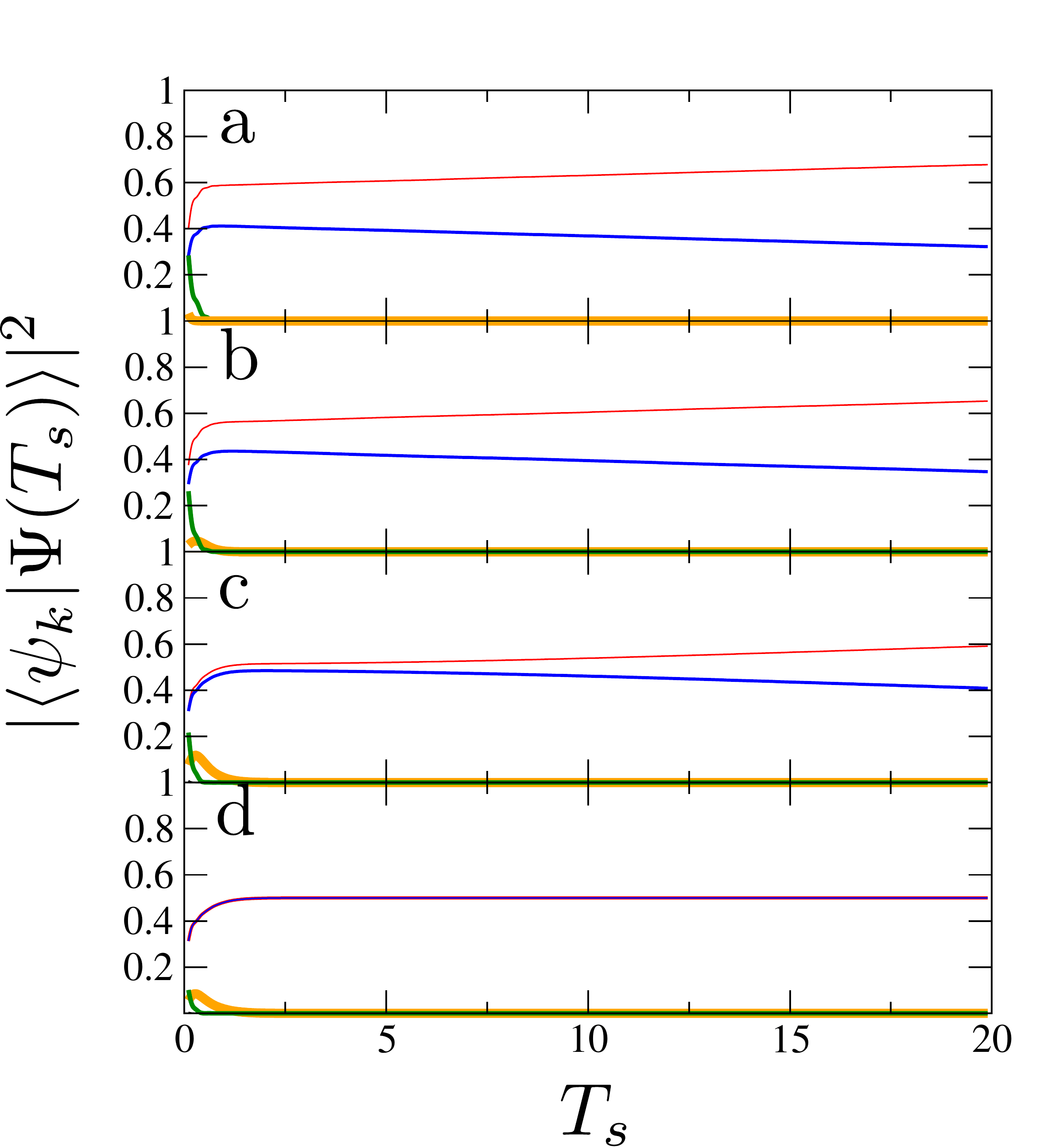}
\caption{\label{fig:switch}(\emph{Color online}) The outcome of the preparation of the wavefunction $\Psi(T_s)$. The plots show the magnitude of the components of $\Psi(t)$ at time $t = T_s$ projected onto the instantaneous eigenbasis $|\psi_k(t)\rangle$ of the full Hamiltonian $H(t)$ (\ref{H0}), as a function of $T_s$ for different fixed values $n_g = 0$ (a), $n_g = 0.5$ (b), $n_g = 0.9$ (c), $n_g = 1$ (d). The different components of the wavefunction $|\langle \psi_k | \Psi(T_s) \rangle|^2$ are indicated by curves in order of increasing thickness, $k = 0$ (red), $k = 1$ (blue), $k = 2$ (green), $k = 3$ (orange). }
\end{figure}

On the other hand, the circumstances under which condition 2) is satisfied follow more complex criteria, and it is necessary to examine the dynamics of the many-body state during the switching interval in more detail. If $E_M$ were set to zero at all times, the instantaneous adiabatic eigenstates of the resulting Hamiltonian $H_0(t)$ are slowly varying, while the adiabatic eigenstates of the full Hamiltonian
\begin{gather}
H(t) = H_0(t) + H_M(t) \ ,
\label{H0}
\end{gather}
accounting for the Majorana coupling $H_M(t) = - i E_M(t) \gamma_2 \gamma_3 \cos \frac{ \phi}{2}$ vary quickly in comparison as $E_M$ is switched on. The wavefunction may be expressed as a coherent superposition of the lowest two eigenstates of $H_0(t)$, which have definite and opposite parity, and evolves under the effective two-level Hamiltonian
\begin{gather}
H_{2\times 2}(t) = - \beta_x(t) \frac{\sigma_x}{2} - \beta_z(t) \frac{\sigma_z}{2}
\label{Hamiltwolevel}
\end{gather}
where $\beta_x(t)$ is the projection of $H_M(t)$ onto the lowest two-level subspace spanned by the eigenstates of $H_0(t)$ and $\beta_z(t)$ is the splitting between these levels. The evolution of the wavefunction is significantly complicated by the nonlinear time dependence of $\beta_z(t) \sim E_J(t)^\frac{3}{4} e^{ - \sqrt{ 2 E_J(t)/E_-}}$ for $t \approx T_s$. If $n_g$ were tuned to a crossing of parity states (see Fig.~\ref{fig:paras}a), so that $\beta_z(t)$ remained zero then the wavefunction would be prepared in a perfect equal superposition of $\sigma_x$ eigenstates for all switching times. However if the detuning of $n_g$ from the crossing were significant,  $\beta_z(t)$ would be of the order of $E_-$ under Coulomb blockade ($t = 0$) and vary exponentially over the switching period. In this situation the evolution of the wavefunction varies considerably as a function of $n_g(t)$ and it is not possible to characterize it via the switching time $T_s$.

This fact is illustrated in the plots in  Figs.~\ref{fig:osc} and \ref{fig:switch} of the time-evolution of the many-body wavefunction $\Psi(t)$, calculated via numerical integration, for a number of protocols, with initialization in the state $|n = 0\rangle$. The plots demonstrate the sensitivity of coherent oscillations to the value of $n_g$ during the switching intervals. In order to simulate the opening of the tunnel barrier, we assume that the tunneling amplitudes, and therefore $E_M$, increase linearly with time, and consequently $E_J$ increases quadratically in time over the initial switching interval.
In the final switching period all parameters undergo the inverse evolution. We choose to keep $n_g$ fixed for the entire process, and this may be contrasted with a typical Ramsey experiment performed in a non-topological Cooper pair box, in which $n_g$ is varied from an initial value at which the ground state exhibits no charge fluctuations to a value corresponding to an anticrossing between charge states, with the Josephson coupling driving coherent oscillations during the waiting period.  \color{black} In our case, it is not necessary to vary $n_g$ since $E_J$ may be used to collapse the spacing between charge states while $E_M$ drives the oscillations. Explicitly, the process is given by
\begin{gather}
E_M(t) = \begin{cases}
\alpha t \ \ , \ \ 0 < t < T_s \\
\alpha T_s \ \ , \ \ T_s < t < T_s + T_w \ \ , \\
\alpha(T_f - t) \ \ , \ \ T_s + T_w < t < T_w
\end{cases} \ \ , \nonumber \\
E_J(t) = \begin{cases}
\beta t^2 \ \ , \ \ 0 < t < T_s \ \ , \\
\beta T_s^2 \ \ , \ \ T_s < t < T_s + T_w  \ \ , \\
\beta (t - T_f)^2 \ \ , \ \ T_s + T_w < t < T_f
\end{cases} \ \ .
\label{protocol}
\end{gather}
In Fig. \ref{fig:osc} we have plotted the full wavefunction for several situations projected onto two different bases: the first (A) consisting of instanteous eigenstates $|\psi_k\rangle $ of the full Hamiltonian $H(t)$ (\ref{H0}), and the second (B) consisting of the instantaneous eigenstates $|\varphi_{l \sigma}\rangle$ of $H_0(t)$ (\ref{H0}) which possess definite parity $\sigma = \pm 1$. Plotted in basis (A) (Fig.~\ref{fig:osc}a, c, e, g, i), the quantum state $\Psi(t)$ evolves from the lowest charge eigenstate into a coherent superposition of the lowest two instantaneous eigenstates of $H(t)$ at the end of the initial switching period, with the weights $\langle \psi_k | \Psi\rangle$ in the states $ k = 0, 1$ depending significantly on the value of $n_g$.  Maximum visibility of oscillations, $|\langle \psi_0 | \Psi\rangle|^2 = |\langle \psi_1 | \Psi\rangle|^2$, is achieved when $n_g = 1$, which corresponds to perfect  initialization in the state $|n = 0\rangle$, although this state becomes degenerate with $|n = 2\rangle$ for this value of $n_g$. \color{black} Significantly lower coherence is achieved for $n_g = 0$, which corresponds to initialization in a ground state which has maximum separation from the higher charge states. Plotted in basis (B) (Fig.~\ref{fig:osc}b,d,f,h,g), the wavefunction exhibits parity oscillations with amplitudes which are only slightly suppressed by imperfect preparation during switching. For $n_g = 0$, the components of the wavefunction in basis (A) are $|\langle \psi_0 | \Psi \rangle |^2 = 0.68$, $|\langle \psi_1 | \Psi \rangle |^2 = 0.32$, and the amplitude of oscillations is $2\sqrt{0.68 \times 0.32} = 0.93$. In all our plots the values of the inter-island coupling constants are $E_M(T_s) = \alpha T_s =  0.05 E_-$, $E_J(T_s) = \alpha T_s^2 = 50 E_-$.

In Fig. \ref{fig:switch} we plot the magnitude of the components of the many-body wavefunction $\langle \psi_k(t) | \Psi(t = T_s)\rangle$ projected onto the instanteous eigenbasis as a function of $T_s$ for various values of $n_g$. Figs. \ref{fig:osc} and \ref{fig:switch}  illustrate the dependence of the evolution of $\Psi(t)$ to the behaviour of $n_g$ during the switching interval. Coherence is maximum when $n_g  = 1$ and decreases substantially as both $n_g$ is driven away from the crossing and $T_s$ is increased. The plots also illustrate the sensitivity of the final conversion process ($T_s + T_w < t < T_f$) to $n_g$: when $n_g$ is moved slightly away from zero, the third level is populated with probability $\approx 0.15$, resulting from the collapse of the level spacing between charge states $|n = 2 \rangle$ and $| n = -2\rangle$ at the end of the process.

\section{Decoherence}

Having established the predicted outcomes of a carefully controlled ideal experiment, we shall now focus on the reduction of the visibility of coherent oscillations due to coupling to a noisy  environment. We may introduce decoherence into our model (\ref{Hamil2}) via random fluctuations of the parameters $E_J(t)$, $E_M(t)$, $n_g(t)$ originating from classical fluctuations of the local electrostatic potential,
which generate random terms in the Hamiltonian
\begin{gather}
H \rightarrow H - \delta n_g(t) V_N - \delta E_J(t) V_J - \delta E_M(t) V_M
\end{gather}
where
\begin{gather}
V_N = - \frac{\partial H}{\partial n_g} = 2 E_- \hat{n} \ \ , \nonumber \\
V_J = \cos \phi \ \ , \ \ V_M = i \gamma_2 \gamma_3 \cos \frac{\phi}{2} \ \ .
\end{gather}
The von Neumann equation for the density matrix in the interaction picture is given by
\begin{gather}
i \frac{d}{dt} \langle \langle \rho^I(t) \rangle \rangle  =  \nonumber \\ \langle \langle \left[ \delta n_g(t) V_N^I(t) + \delta E_J(t) V_J^I(t) + \delta E_M(t) V_M^I(t), \rho^I(t)\right]
\rangle \rangle
\end{gather}
with $\rho^I(t) = U^\dagger(t) \rho(t) U(t)$, $V_N^I(t)= U^\dagger(t) V_N U(t)$, $V_J^I(t) = U^\dagger(t) V_J U(t)$, $V_M^I(t) = U^\dagger(t) V_M U(t)$ and  $U$ being the propagator unperturbed by random fluctuations. If we assume that the correction to the measurement outcome $\text{Tr} P_0\rho(T_f)$ (\ref{measurement}) is small, we may expand the density matrix to lowest order in the autocorrelation functions of the fluctuations, and the resulting correction to $\text{Tr}  P_0 \rho(T_f)$ consists of a sum of separate contributions from fluctuations in the three parameters,
\begin{gather}
\delta \text{Tr} P_0 \rho(T_f) =- \text{Tr} P_0 \left[ \delta_V+ \delta_J + \delta_M \right]
\end{gather}
where
\begin{align}
\delta_V =&
\int_0^{T_f}\!\!\! \int_0^t \langle\langle \delta n_g(t) \delta n_g(t') \rangle \rangle \notag\\ &\times\left[ V^I_N(t), \left[ V^I_N(t'), \rho(0)\right]\right] dt'dt,
\label{doubleintV}\\
\delta_J =&
\int_0^{T_f} \int_0^t \langle \langle \delta E_J(t) \delta E_J(t') \rangle \rangle\notag\\ &\times \left[ V^I_J(t), \left[ V^I_J(t'), \rho(0)\right]\right] dt'dt,
\label{doubleintJ}\\
\delta_M =&
\int_0^{T_f} \int_0^t \langle \langle \delta E_M(t) \delta E_M(t') \rangle \rangle\notag\\ &\times \left[ V_M^I(t), \left[ V_M^I(t'), \rho(0)\right]\right] dt'dt.
\label{doubleintM}
\end{align}
As we saw in the previous section, the effect of the unperturbed propagator $U$ is to generate rotations within each two-dimensional subspace spanned by pairs of instantaneous eigenstates $(\varphi_{l +}, \varphi_{l-})$ of $H_0(t)$ (\ref{H0}). Low-frequency fluctuations in $\delta n_g(t)$, $\delta E_M(t)$, $\delta E_J(t)$, enter the integrals (\ref{doubleintV},\ref{doubleintJ},\ref{doubleintM}) via the matrix elements of the operators $V_N$, $V_J$ and $V_M$ among states lying in the lowest two-level subspace, while higher frequency fluctuations may result in the generation of nonvanishing components of the density matrix in the higher levels, which we will discuss in detail later. Considering the influence of low-frequency fluctuations, we find that $\langle \varphi_{l\sigma} | V_J | \varphi_{l \sigma'} \rangle = \langle \varphi_{l \sigma} | V_M | \varphi_{l \sigma} \rangle =0$ while $\langle \varphi_{l+} | V_M | \varphi_{l_-}\rangle$ varies between unity at $t = 0$ (when $E_J = 0$) and two at $t = T_s$ (when $E_J \gg E_-$). It follows that $\delta_J$ vanishes, while fluctuations in $E_M$ lead to pure dephasing, with
\begin{gather}
-\text{Tr} P_0 \delta_M \approx 4\int_{0}^{T_f}{ \int_{0}^{t}{\langle \langle \delta E_M(t) \delta E_M(t') \rangle \rangle dt'}dt} \ \ .
\end{gather}
The value of $E_M$ is controlled by the transparency of the central junction, and we should expect  $\delta E_M$ to vary significantly during the switching periods.

For sufficiently large values of $E_J/E_-$, the contribution to decoherence from low-frequency fluctuations in $n_g$ becomes exponentially suppressed since the spectrum of $H_0(t)$ is insensitive to $n_g$, which implies that $\langle \varphi_{l\sigma} | V_N | \varphi_{l \sigma'} \rangle = \langle \varphi_{l \sigma} | - \frac{ \partial H}{\partial n_g} | \varphi_{l \sigma'}\rangle \approx 0$. We may then evaluate (\ref{doubleintV}) by expressing $U(t)$ as a rotation matrix within the subspace $(\varphi_{l +}, \varphi_{l -})$ with $l = 0$ via $U^\dagger(t) \bm{\sigma} U(t) = \bm{\sigma} R(t)$ to obtain
\color{black}
\begin{gather}
\delta \text{Tr} \rho(T_f) \sigma_z = -(2 E_-)^2\times \nonumber \\
\int_0^{T_f}{ \int_0^{t} {
		\langle \langle \delta n_g(t) \delta n_g(t') \rangle \rangle u(t) u(t')  R(T_f)\hat{z}\cdot \mathcal{Z}(t',t) dt'dt }} \ \ , \nonumber \\
\mathcal{Z}(t',t) =   R(t) \hat{z} \times (R(t')\hat{z}\times \hat{z}) \ \ ,
\label{doubleintR}
\end{gather}
where
\begin{gather}
u(t) =\frac{  \langle \varphi_{0+} | \hat{n}  | \varphi_{0+} \rangle - \langle  \varphi_{0-} | \hat{n} |  \varphi_{0 - }\rangle}{2}  \ \ .
\end{gather}
Noting that $u(t)$ is equal to unity at $t = 0$ and becomes suppressed by the large value of $E_J$ for $T_s < t < T_s + T_w$, we only need to consider contributions to the integral (\ref{doubleintR}) from times $t,t'$ both lying within the switching periods. We also note that for most of the initial switching period $\beta_z \gg \beta_x$ and therefore $R(t)$ generates rotations about the $\hat{z}$ axis, and the factor $R(t') \hat{z} \times \hat{z} = 0$. Consistent with these observations, numerical evaluation of (\ref{doubleintR}) shows that the product of the factors $u(t), u(t')$ and those arising from rotations is negligible for $t, t'< T_s$. On the other hand, at the end of the waiting period, $t = T_s + T_w$, the polarization vector $\text{Tr} \bm{\sigma} \rho(t)$ lies in the $y-z$ plane and may undergo significant rotation during the final stage of the protocol, $T_s + T_w < t' < t < T_f$. We therefore only account for the contributions to the integral from the final stage.

Introducing the rotation operators
\begin{gather}
R_1 = R(T_s) \ \ , \ \ R_2 = R(T_s + T_w)R^\dagger(T_s) \ \ , \nonumber \\
R_3(t) = R(t) R^\dagger(T_s + T_w) \ \ ,
\end{gather}
and noting that over the waiting period $T_s < t, t'< T_s + T_w$ the two-level Hamiltonian (\ref{Hamiltwolevel}) is proportional to $\sigma_x$, we observe that $R_2$ is simply a rotation about the $\hat{x}$-axis by an angle $\omega T_w$ where $\omega = \beta_z(T_s)$. Thus we may express the vectors $R(t)\hat{z}$ for $T_s + T_w < t < T_f$ in terms of the waiting time $T_s$ via
\begin{gather}
R(t) = R_3(t) R_2 R_1 \hat{z}\ \ , \nonumber \\
R_1 \hat{z} = v_0 \hat{n}_0 + \frac{ v_1 \hat{n}_{-1} + v_{-1} \hat{n}_1 }{2} \ \ , \nonumber \\
R_2 R_1 \hat{z} = v_0 \hat{n}_0 + \frac{v_1 e^{-i\omega T_w} \hat{n}_{-1} + v_{-1} e^{i \omega T_w} \hat{n}_{1}}{2}
\end{gather}
where
\begin{gather}
\hat{n}_0 = \hat{x} \ \ , \ \
\hat{n}_{\pm 1} = \hat{z}\mp i \hat{y} \ \  .
\end{gather}
In terms of the quantities associated with the rotation operators the measurement outcome in the absence of decoherence is given by
\begin{gather}
\langle P_0 \rangle_{\text{final}} 
= \frac{1}{2} \left[ 1 +   \lambda_0 + \lambda_1 \cos( \omega T_w + \Phi) \right] \ \ , \nonumber \\
\lambda_0 = (\hat{x} \cdot R_1 \hat{z}) ( \hat{z} \cdot R_3(T_f) \hat{x}) \ \ , \nonumber \\
\lambda_1 = \sqrt{ 1 - (\hat{x} \cdot R_1 \hat{z})^2} \sqrt{1 - ( \hat{z}\cdot R_3(T_f) \hat{x})^2} \ \ , \nonumber \\
\Phi =\tan^{-1} \frac{\hat{y} \cdot R_1 \hat{z}}{ \hat{z} \cdot R_1 \hat{z}} - \tan^{-1} \frac{\hat{z} \cdot R_3(T_f) \hat{y}}{ \hat{z} \cdot R_3(T_f) \hat{z}} \ \ .
\end{gather}
Since the integrand in (\ref{doubleintR}) contains products of the vectors $R_3(t)$, $R_3(t' )$ and $R_3(T_f)$, the correction to $\langle P_0 \rangle_{\text{final}}$ due to decoherence takes the form of a sum over harmonics
\begin{gather}
-\delta \langle P_0 \rangle_{\text{final}}= 
\frac{1}{2} W_0 +\frac{1}{2} \text{Re} \left[ W_1 e^{i \omega T_w} + W_2 e^{2i \omega T_w} + W_3 e^{3i \omega T_w} \right] \ \ ,
\end{gather}
where
\begin{widetext}
\begin{equation}\label{conversionerror}
\begin{aligned}
W_n =& -\!\!\!\!\! \sum_{i+j+k = -n}  v_i v_j v_k\int_{T_s + T_w}^{T_f} \int_{T_s + T_w}^{t}
\langle \langle \delta n_g(t) \delta n_g(t') \rangle \rangle u(t) u(t') R_3(T_f)\hat{n}_i \cdot \left[  R_3(t) \hat{n}_j \times (R_3(t') \hat{n}_k \times \hat{z})\right]  dt' dt 
\end{aligned}
\end{equation}
\end{widetext}
\color{black}
In the case of ideal preparation during the switching interval, the polarization $\text{Tr} \rho(t) \bm{\sigma}$ lies in the $y-z$ plane and undergoes rotations about the $x$-axis. Consequently $v_0$ vanishes and $|v_1| = |v_{-1}| = 1$. We therefore expect that $W_n$ are small for $n = 0, 2$. Generally, the rotational factors as well as $u(t)$ in (\ref{Hamiltwolevel}) are of the order of unity in the relevant domains, and if significant correlations exist during the final switching period, $\langle \langle \delta n_g(t) \delta n_g(t') \rangle \rangle \sim \langle \delta n_g^2 \rangle$ where $\sqrt{ \langle \delta n_g^2 \rangle }$ is the rms value of the fluctuations in $n_g$. The maximum possible value of the corrections due to decoherence for $n = 1, 3$  is then
\begin{gather}
|W_n | \rightarrow \langle \delta n_g^2 \rangle (2 E_-T_s)^2 \ \ .
\end{gather}

\begin{figure}[t!]
	\includegraphics[width = 0.5\textwidth]{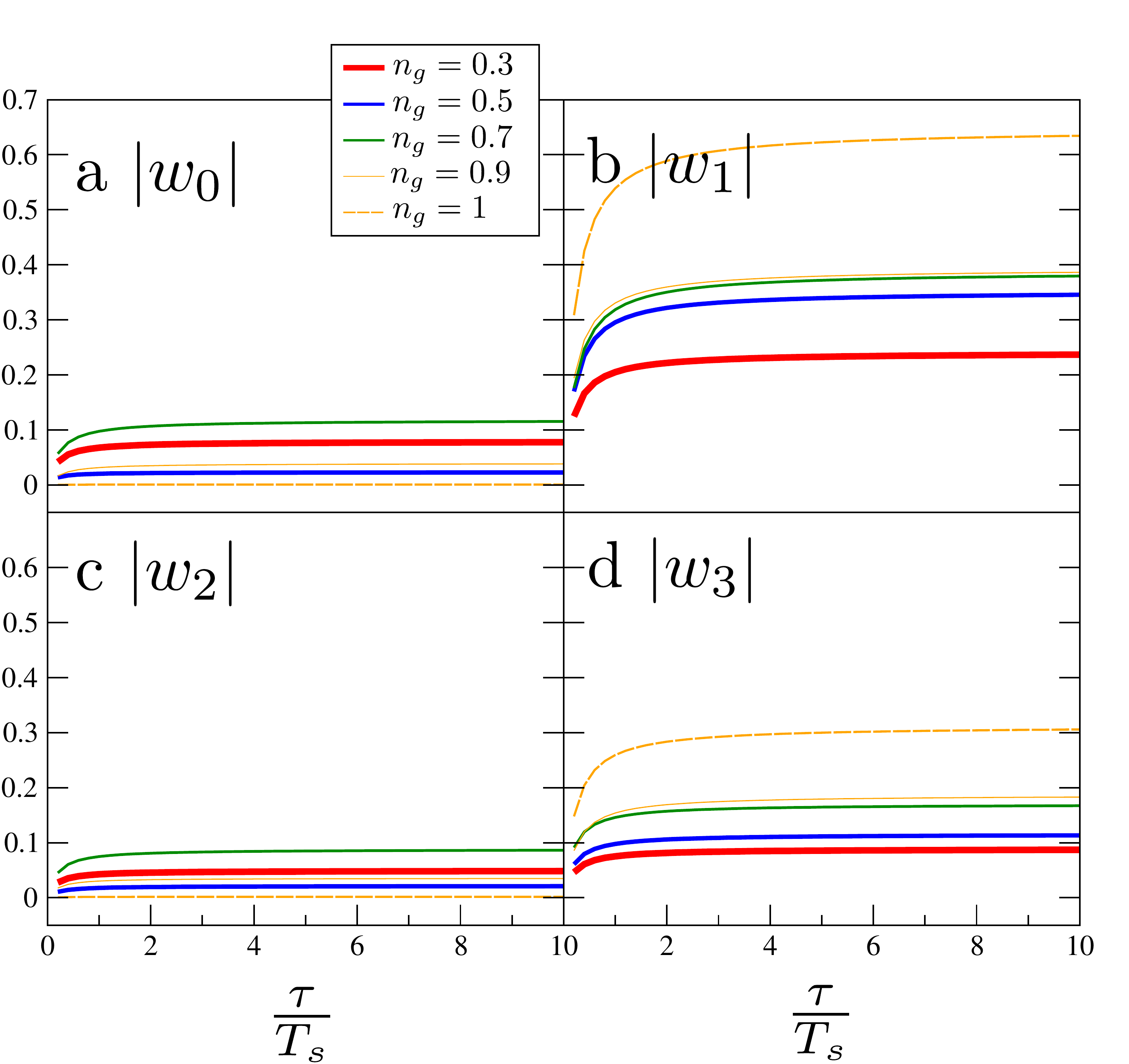}
	\caption{\label{fig:W}(\emph{Color online}) The magnitude of coefficients $|w_n| = |W_n|/\langle \delta n_g^2 \rangle (2 E_- T_s)^2$ [Eq. \ref{conversionerror}], plotted in curves in order of decreasing thickness $n_g = 0.3$ (red), 0.5 (blue), 0.7 (green), and 0.9 (orange). The dashed orange curve shows $n_g =  1$.}
\end{figure}

\begin{figure*}[t!]
	\includegraphics[width = \textwidth]{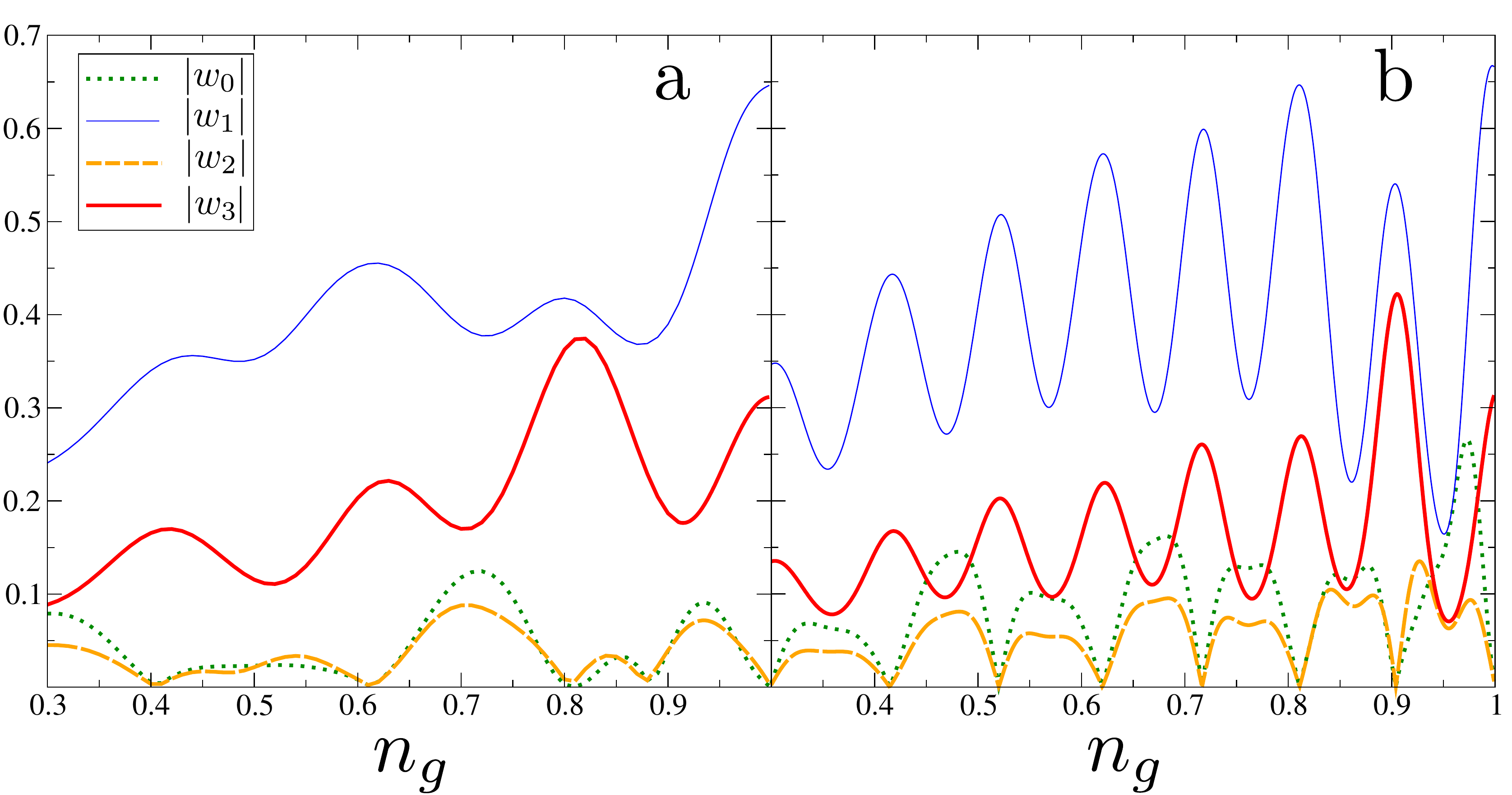}
	\caption{\label{fig:Wsat}(\emph{Color online}) The saturating values of $|w_n|$ as a function of $n_g$ for two protocols with $T_s = 10/E_-$ (left) and $T_s = 20/E_-$ (right).}
\end{figure*}

We evaluated the integrals (\ref{conversionerror}) numerically with an autocorrelation function of the form
\begin{gather}
\langle \langle \delta n_g(t') \delta n_g(t) \rangle\rangle =\langle \delta n_g^2 \rangle e^{- |t - t'|/\tau}
\end{gather}
which corresponds to a single Lorentzian fluctuator with a correlation time $\tau$.
The protocol is given in Eq. (\ref{protocol}), with the coupling constants $E_J = 50 E_-$, $E_M = 0.05 E_-$ during the waiting period, and certain fixed values of $n_g$. We parametrize the correction via dimensionless constants $|w_n|$ defined by
\begin{gather}
|W_n| = |w_n| \langle \delta n_g^2 \rangle (2E_- T_s)^2 \ \ .
\end{gather}
We have plotted  $|w_n|$ for the case where the switching time is $T_s = 10/E_-$ in Fig.~\ref{fig:W} as a function of the correlation time for values $n_g = 0.3$ (red), 0.5 (blue),  0.7 (green), 0.9 (orange, dashed), and 1 (orange). The plots exhibit saturation for $\tau \gg T_s$, with the influence of decoherence being maximum when $n_g = 1$, in which case $|w_1| \approx 0.6$ and $|w_3| \approx 0.3$. We have plotted the dependence of the saturating values of $|w_n|$ on $n_g$ for $T_s = 10/E_-$ and $T_s = 20/ E_-$ in panels a, b respectively of Fig.~\ref{fig:Wsat}. When $|w_1|, |w_3|$ are of the order of unity, the correction to the oscillations is of the order of $\langle 400 \delta n_g^2 \rangle$ for $T_s =10/E_-$, and the  oscillations are significantly reduced when the rms value of the fluctuations exceeds $5\%$. Both plots display a highly monotonic dependence of the decoherence corrections to the visibility on $n_g$, as well an extreme sensitivity, with $|w_1|$ varying by $\approx 50\%$ over the range $n_g = 0.9$ and $n_g = 1$ in both panels of Fig.~\ref{fig:Wsat}.

We also note that for $n_g \approx 0$, leakage outside of the operational space occurs at the end of the protocol, as was shown in Fig.~\ref{fig:osc}. In such a situation our formalism based on Eq. \ref{conversionerror} is no longer applicable, thus we only show results in the range $0.3 \leq n_g < 1$.

\twocolumngrid

In addition to decoherence arising from rotations of the density matrix during the final stage of the protocol, we must also consider contributions to the integral (\ref{doubleintV}) containing matrix elements of the perturbing operator $V$ which connect the operational subspace to the higher levels. At the beginning and end of the protocol, the instantaneous spectrum of the Hamiltonian consists of states of definite charge and $V$ becomes diagonal, so the integral (\ref{doubleintV}) does not contain contributions from times where $E_J \ll E_-$. Considering the situation where the increase in $E_J$ is most rapid at the end of the initial switching period and the beginning of the final switching period (e.g. $E_J \sim t^2$ and $E_J \sim (T_f - t)^2$), we may take the limits of integration in (\ref{doubleintV}) to enclose the waiting period, during which the instantaneous spectrum of the complete Hamiltonian $H(t)$ consists of equal superpositions of oscillator states with opposite parity. The operator $V = -8i E_- \frac{d}{d\phi}$
is a sum of creation and annihilation operators acting on the oscillator levels, which couples the lowest four eigenstates $\psi_k$ of $H$ pairwise, with matrix elements
\begin{gather}
\langle \psi_2 | V | \psi_0 \rangle = \langle \psi_3 | V | \psi_1 \rangle = \sqrt{ \omega_p E_-} \ \ .
\end{gather}
Fermi's golden rule then implies that an incoherent population of the higher levels spanned by $(\psi_2, \psi_3)$ will occur at a rate proportional to the component of the noise spectrum at the plasma frequency. Explicitly evaluating Eq.~(\ref{doubleintV}) for a general spectrum, $\langle \langle \delta n_g(t) \delta n_g(0 ) \rangle \rangle = \int{ S(\omega) e^{-i \omega t} d\omega/2\pi}$ we obtain
\begin{gather}
\frac{\delta \langle P_0 \rangle}{\langle P_0 \rangle} = - E_-\omega_p  (S (\omega_p ) T_w +  \nonumber \\
\text{Re} \left[ e^{-i \omega_p T_w} \int{ \frac{S(\omega) e^{i \omega T_w}}{(\omega - \omega_p + i0)^2} \frac{d\omega}{2\pi}}
 - \int{ \frac{S(\omega)}{ ( \omega - \omega_p + i0)^2} \frac{d\omega}{2\pi}}\right]
\ \ .
\end{gather}
In addition to the linear decay, there is a correction to the visibility which oscillates  at the plasma frequency as a function of $T_w$. For realistic parameters $E_- \approx 10~\mu\text{eV}$, $E_J/E_- = 50$, the plasma frequency is $\approx 600 \text{ GHz}$, and the oscillating contribution is completely invisible. Furthermore, there is no contribution from low-frequency noise, and the only relevant source of potential fluctuations consists of high-frequency Johnson-Nyquist noise. In this range, $S(\omega)$ is independent of frequency and linear decay is enhanced with increasing $E_J$ due to the fact that the probability for excitations between oscillator levels is proportional to the plasma frequency.

\section{Summary and conclusions}

In this work we have presented a detailed study of a four-Majorana qubit which, despite its simple layout, may provide key insight into the degree of coherent control over Majorana couplings which is essential to future devices for Majorana-based quantum computation. Our analysis demonstrates that the system may be operated smoothly between the charge qubit and transmon regimes while retaining a two-level operational subspace corresponding to the parity degree of freedom which is inherited from the topological superconductivity of the islands. We also show how charge sensing in equilibrium may be used to measure both $E_J$ and $E_M$ in the physically interesting regimes.

We have investigated a protocol in which the qubit is initialized and read out in the charge basis but operated as a transmon qubit while coherent oscillations of the parity are performed.  A novel feature of our qubit is that the inverse qubit frequency $  \frac{2\pi}{E_M}$ may be made long while maintaining $E_J \gg E_-$, in comparison to conventional transmons for which the inverse frequency $ \frac{2\pi}{\omega_p} = \frac{\pi}{ \sqrt{32 E_J E_-}}$ must decrease as the system is driven deeper into the large-$E_J$ regime. However, choosing to initialize and read out in the charge basis introduces significant errors which accumulate over the final switching interval when a fluctuating electrostatic environment is present, and this provides a major obstacle to performing charge-based readout which we expect will be shared by other schemes based on parity-to-charge conversion (see, for example Refs. \cite{Aasen2016, Hell2016}).

We have considered several mechanisms for the decoherence of the device during coherent oscillations, namely 1) conversion errors resulting from $\delta n_g$ fluctuations, 2) pure dephasing resulting from $\delta E_M$ fluctuations and 3) leakage out of the operational space due to high-frequency $\delta V_g$ fluctuations during the waiting period.

We have shown that conversion errors lead to a correction to the oscillations which is generally $\sim \langle 2 E_- \delta n_g^2\rangle T_s^2 = \frac{1}{4} \langle \delta V_g^2 \rangle T_s^2$, which is comparable to decoherence in conventional charge qubits due to the same source of noise. This particular mechanism decoheres the qubit only during the final switching interval and therefore influences the shape of the oscillations via the introduction of other harmonics without affecting their visibility at long wait times. Based on detailed analysis, we have shown that this mechanism as well as the qubit initialization are highly sensitive to the value of $n_g$: when $n_g = 1$, corresponding to initialization at a level crossing of charge states, preparation is ideal in the absence of noise, with the many-body wavefunction having probabilities $|\langle \psi_0|\Psi \rangle|^2 = | \langle \psi_1 | \Psi\rangle|^2 = 0.5$ in the lowest two levels $\psi_0, \psi_1$ of the instantaneous eigenbasis during the waiting period, regardless of the switching time. At the same time, the qubit is most susceptible to noise during the final switching interval. When $n_g = 0.5$, the decoherence correction to the oscillations is reduced (Fig. \ref{fig:Wsat}), however the preparation is far from ideal, with $|\langle \psi_0 | \Psi\rangle|^2 = 0.65$ and $|\langle\psi_1 | \Psi\rangle|^2 = 0.35$ for $T_s = 20/E_-$ (Fig. \ref{fig:switch}b). We note that if the protocol is altered to introduce time-variation in $n_g$, then both initialization and conversion errors will depend only weakly on the value of $n_g$ outside the waiting period, as the qubit state undergoes the maximum rotation at times $t \approx T_s, T_w + T_s$. 

In contrast, both $\delta E_M$ and $\delta V_g$ fluctuations during the waiting period lead to linear decay of the oscillations at short times, with $\delta E_M$ associated with pure dephasing and $\delta V_g$ with leakage into the excited states of the transmon spectrum when the noise spectrum possesses weight at the plasma frequency.

The coexistence of several mechanisms of decoherence which have qualitatively distinct influences on the oscillations demonstrates that a significant amount of information about the interaction of the fluctuating electrostatic environment with the dynamics of the many-body wavefunction may be gleaned from the presence of higher harmonics in the oscillations. All in all our findings provide useful tools for studying future Majorana-based qubit designs which are susceptible to the same mechanisms of decoherence when operated outside the topologically protected regime.

\section{Acknowledgments}

The authors acknowledge financial support from the Crafoord Foundation and the Swedish Research Council. W.A.C. acknowledges support from NSERC, CIFAR, FRQNT, Nordea Fonden, and the Gordon Godfrey Bequest. This work was supported by the Danish National Research Foundation and by the Microsoft Station Q Program.

\bibliography{Majorana}

\begin{thebibliography}{36}%
\makeatletter
\providecommand \@ifxundefined [1]{%
 \@ifx{#1\undefined}
}%
\providecommand \@ifnum [1]{%
 \ifnum #1\expandafter \@firstoftwo
 \else \expandafter \@secondoftwo
 \fi
}%
\providecommand \@ifx [1]{%
 \ifx #1\expandafter \@firstoftwo
 \else \expandafter \@secondoftwo
 \fi
}%
\providecommand \natexlab [1]{#1}%
\providecommand \enquote  [1]{``#1''}%
\providecommand \bibnamefont  [1]{#1}%
\providecommand \bibfnamefont [1]{#1}%
\providecommand \citenamefont [1]{#1}%
\providecommand \href@noop [0]{\@secondoftwo}%
\providecommand \href [0]{\begingroup \@sanitize@url \@href}%
\providecommand \@href[1]{\@@startlink{#1}\@@href}%
\providecommand \@@href[1]{\endgroup#1\@@endlink}%
\providecommand \@sanitize@url [0]{\catcode `\\12\catcode `\$12\catcode
  `\&12\catcode `\#12\catcode `\^12\catcode `\_12\catcode `\%12\relax}%
\providecommand \@@startlink[1]{}%
\providecommand \@@endlink[0]{}%
\providecommand \url  [0]{\begingroup\@sanitize@url \@url }%
\providecommand \@url [1]{\endgroup\@href {#1}{\urlprefix }}%
\providecommand \urlprefix  [0]{URL }%
\providecommand \Eprint [0]{\href }%
\providecommand \doibase [0]{http://dx.doi.org/}%
\providecommand \selectlanguage [0]{\@gobble}%
\providecommand \bibinfo  [0]{\@secondoftwo}%
\providecommand \bibfield  [0]{\@secondoftwo}%
\providecommand \translation [1]{[#1]}%
\providecommand \BibitemOpen [0]{}%
\providecommand \bibitemStop [0]{}%
\providecommand \bibitemNoStop [0]{.\EOS\space}%
\providecommand \EOS [0]{\spacefactor3000\relax}%
\providecommand \BibitemShut  [1]{\csname bibitem#1\endcsname}%
\let\auto@bib@innerbib\@empty
\bibitem [{\citenamefont {Read}\ and\ \citenamefont {Green}(2000)}]{Read2000}%
  \BibitemOpen
  \bibfield  {author} {\bibinfo {author} {\bibfnamefont {N.}~\bibnamefont
  {Read}}\ and\ \bibinfo {author} {\bibfnamefont {D.}~\bibnamefont {Green}},\
  }\href {\doibase 10.1103/PhysRevB.61.10267} {\bibfield  {journal} {\bibinfo
  {journal} {Phys.~Rev.~B}\ }\textbf {\bibinfo {volume} {61}},\ \bibinfo
  {pages} {10267} (\bibinfo {year} {2000})}\BibitemShut {NoStop}%
\bibitem [{\citenamefont {Volovik}\ and\ \citenamefont
  {Yakovenko}(1989)}]{Volovik1989}%
  \BibitemOpen
  \bibfield  {author} {\bibinfo {author} {\bibfnamefont {G.~E.}\ \bibnamefont
  {Volovik}}\ and\ \bibinfo {author} {\bibfnamefont {V.~M.}\ \bibnamefont
  {Yakovenko}},\ }\href {\doibase 10.1088/0953-8984/1/31/025} {\bibfield
  {journal} {\bibinfo  {journal} {J.~Phys.:~Condens.~Matter}\ }\textbf
  {\bibinfo {volume} {1}},\ \bibinfo {pages} {5263} (\bibinfo {year}
  {1989})}\BibitemShut {NoStop}%
\bibitem [{\citenamefont {Kitaev}(2001)}]{Kitaev2001}%
  \BibitemOpen
  \bibfield  {author} {\bibinfo {author} {\bibfnamefont {A.~Y.}\ \bibnamefont
  {Kitaev}},\ }\href {\doibase 10.1070/1063-7869/44/10S/S29} {\bibfield
  {journal} {\bibinfo  {journal} {Phys.~Usp.}\ }\textbf {\bibinfo {volume}
  {44}},\ \bibinfo {pages} {131} (\bibinfo {year} {2001})}\BibitemShut
  {NoStop}%
\bibitem [{\citenamefont {Nayak}\ \emph {et~al.}(2008)\citenamefont {Nayak},
  \citenamefont {Simon}, \citenamefont {Stern}, \citenamefont {Freedman},\ and\
  \citenamefont {{Das Sarma}}}]{Nayak2008}%
  \BibitemOpen
  \bibfield  {author} {\bibinfo {author} {\bibfnamefont {C.}~\bibnamefont
  {Nayak}}, \bibinfo {author} {\bibfnamefont {S.~H.}\ \bibnamefont {Simon}},
  \bibinfo {author} {\bibfnamefont {A.}~\bibnamefont {Stern}}, \bibinfo
  {author} {\bibfnamefont {M.}~\bibnamefont {Freedman}}, \ and\ \bibinfo
  {author} {\bibfnamefont {S.}~\bibnamefont {{Das Sarma}}},\ }\href {\doibase
  10.1103/RevModPhys.80.1083} {\bibfield  {journal} {\bibinfo  {journal}
  {Rev.~Mod.~Phys.}\ }\textbf {\bibinfo {volume} {80}},\ \bibinfo {pages}
  {1083} (\bibinfo {year} {2008})}\BibitemShut {NoStop}%
\bibitem [{\citenamefont {Alicea}(2012)}]{AliceaReview}%
  \BibitemOpen
  \bibfield  {author} {\bibinfo {author} {\bibfnamefont {J.}~\bibnamefont
  {Alicea}},\ }\href {\doibase 10.1088/0034-4885/75/7/076501} {\bibfield
  {journal} {\bibinfo  {journal} {Rep. Prog. Phys.}\ }\textbf {\bibinfo
  {volume} {75}},\ \bibinfo {pages} {076501} (\bibinfo {year}
  {2012})}\BibitemShut {NoStop}%
\bibitem [{\citenamefont {Leijnse}\ and\ \citenamefont
  {Flensberg}(2012)}]{LeijnseReview}%
  \BibitemOpen
  \bibfield  {author} {\bibinfo {author} {\bibfnamefont {M.}~\bibnamefont
  {Leijnse}}\ and\ \bibinfo {author} {\bibfnamefont {K.}~\bibnamefont
  {Flensberg}},\ }\href {\doibase 10.1088/0268-1242/27/12/124003} {\bibfield
  {journal} {\bibinfo  {journal} {Semicond.~Sci.~Technol.}\ }\textbf {\bibinfo
  {volume} {27}},\ \bibinfo {pages} {124003} (\bibinfo {year}
  {2012})}\BibitemShut {NoStop}%
\bibitem [{\citenamefont {Beenakker}(2013)}]{BeenakkerReview}%
  \BibitemOpen
  \bibfield  {author} {\bibinfo {author} {\bibfnamefont {C.~W.~J.}\
  \bibnamefont {Beenakker}},\ }\href {\doibase
  10.1146/annurev-conmatphys-030212-184337} {\bibfield  {journal} {\bibinfo
  {journal} {Annu. Rev. Condens. Matter Phys.}\ }\textbf {\bibinfo {volume}
  {4}},\ \bibinfo {pages} {113} (\bibinfo {year} {2013})}\BibitemShut {NoStop}%
\bibitem [{\citenamefont {Fu}\ and\ \citenamefont {Kane}(2008)}]{Fu2008}%
  \BibitemOpen
  \bibfield  {author} {\bibinfo {author} {\bibfnamefont {L.}~\bibnamefont
  {Fu}}\ and\ \bibinfo {author} {\bibfnamefont {C.~L.}\ \bibnamefont {Kane}},\
  }\href {\doibase 10.1103/PhysRevLett.100.096407} {\bibfield  {journal}
  {\bibinfo  {journal} {Phys.~Rev.~Lett.}\ }\textbf {\bibinfo {volume} {100}},\
  \bibinfo {pages} {096407} (\bibinfo {year} {2008})}\BibitemShut {NoStop}%
\bibitem [{\citenamefont {Sau}\ \emph {et~al.}(2010)\citenamefont {Sau},
  \citenamefont {Tewari}, \citenamefont {Lutchyn}, \citenamefont {Stanescu},\
  and\ \citenamefont {{Das Sarma}}}]{Sau2010b}%
  \BibitemOpen
  \bibfield  {author} {\bibinfo {author} {\bibfnamefont {J.}~\bibnamefont
  {Sau}}, \bibinfo {author} {\bibfnamefont {S.}~\bibnamefont {Tewari}},
  \bibinfo {author} {\bibfnamefont {R.}~\bibnamefont {Lutchyn}}, \bibinfo
  {author} {\bibfnamefont {T.}~\bibnamefont {Stanescu}}, \ and\ \bibinfo
  {author} {\bibfnamefont {S.}~\bibnamefont {{Das Sarma}}},\ }\href {\doibase
  10.1103/PhysRevB.82.214509} {\bibfield  {journal} {\bibinfo  {journal}
  {Phys.~Rev.~B}\ }\textbf {\bibinfo {volume} {82}},\ \bibinfo {pages} {214509}
  (\bibinfo {year} {2010})}\BibitemShut {NoStop}%
\bibitem [{\citenamefont {Alicea}(2010)}]{Alicea2010}%
  \BibitemOpen
  \bibfield  {author} {\bibinfo {author} {\bibfnamefont {J.}~\bibnamefont
  {Alicea}},\ }\href {\doibase 10.1103/PhysRevB.81.125318} {\bibfield
  {journal} {\bibinfo  {journal} {Phys.~Rev.~B}\ }\textbf {\bibinfo {volume}
  {81}},\ \bibinfo {pages} {125318} (\bibinfo {year} {2010})}\BibitemShut
  {NoStop}%
\bibitem [{\citenamefont {Lutchyn}\ \emph {et~al.}(2010)\citenamefont
  {Lutchyn}, \citenamefont {Sau},\ and\ \citenamefont
  {Das~Sarma}}]{Lutchyn2010}%
  \BibitemOpen
  \bibfield  {author} {\bibinfo {author} {\bibfnamefont {R.~M.}\ \bibnamefont
  {Lutchyn}}, \bibinfo {author} {\bibfnamefont {J.~D.}\ \bibnamefont {Sau}}, \
  and\ \bibinfo {author} {\bibfnamefont {S.}~\bibnamefont {Das~Sarma}},\ }\href
  {\doibase 10.1103/PhysRevLett.105.077001} {\bibfield  {journal} {\bibinfo
  {journal} {Phys. Rev. Lett.}\ }\textbf {\bibinfo {volume} {105}},\ \bibinfo
  {pages} {077001} (\bibinfo {year} {2010})}\BibitemShut {NoStop}%
\bibitem [{\citenamefont {Oreg}\ \emph {et~al.}(2010)\citenamefont {Oreg},
  \citenamefont {Refael},\ and\ \citenamefont {von Oppen}}]{Oreg2010}%
  \BibitemOpen
  \bibfield  {author} {\bibinfo {author} {\bibfnamefont {Y.}~\bibnamefont
  {Oreg}}, \bibinfo {author} {\bibfnamefont {G.}~\bibnamefont {Refael}}, \ and\
  \bibinfo {author} {\bibfnamefont {F.}~\bibnamefont {von Oppen}},\ }\href
  {\doibase 10.1103/PhysRevLett.105.177002} {\bibfield  {journal} {\bibinfo
  {journal} {Phys.~Rev.~Lett.}\ }\textbf {\bibinfo {volume} {105}},\ \bibinfo
  {pages} {177002} (\bibinfo {year} {2010})}\BibitemShut {NoStop}%
\bibitem [{\citenamefont {Das}\ \emph {et~al.}(2012)\citenamefont {Das},
  \citenamefont {Ronen}, \citenamefont {Most}, \citenamefont {Oreg},
  \citenamefont {Heiblum},\ and\ \citenamefont {Shtrikman}}]{Das2012}%
  \BibitemOpen
  \bibfield  {author} {\bibinfo {author} {\bibfnamefont {A.}~\bibnamefont
  {Das}}, \bibinfo {author} {\bibfnamefont {Y.}~\bibnamefont {Ronen}}, \bibinfo
  {author} {\bibfnamefont {Y.}~\bibnamefont {Most}}, \bibinfo {author}
  {\bibfnamefont {Y.}~\bibnamefont {Oreg}}, \bibinfo {author} {\bibfnamefont
  {M.}~\bibnamefont {Heiblum}}, \ and\ \bibinfo {author} {\bibfnamefont
  {H.}~\bibnamefont {Shtrikman}},\ }\href {\doibase 10.1038/nphys2479}
  {\bibfield  {journal} {\bibinfo  {journal} {Nat.~Phys.}\ }\textbf {\bibinfo
  {volume} {8}},\ \bibinfo {pages} {887} (\bibinfo {year} {2012})}\BibitemShut
  {NoStop}%
\bibitem [{\citenamefont {Mourik}\ \emph {et~al.}(2012)\citenamefont {Mourik},
  \citenamefont {Zuo}, \citenamefont {Frolov}, \citenamefont {Plissard},
  \citenamefont {Bakkers},\ and\ \citenamefont {Kouwenhoven}}]{Mourik2012}%
  \BibitemOpen
  \bibfield  {author} {\bibinfo {author} {\bibfnamefont {V.}~\bibnamefont
  {Mourik}}, \bibinfo {author} {\bibfnamefont {K.}~\bibnamefont {Zuo}},
  \bibinfo {author} {\bibfnamefont {S.~M.}\ \bibnamefont {Frolov}}, \bibinfo
  {author} {\bibfnamefont {S.~R.}\ \bibnamefont {Plissard}}, \bibinfo {author}
  {\bibfnamefont {E.~P. A.~M.}\ \bibnamefont {Bakkers}}, \ and\ \bibinfo
  {author} {\bibfnamefont {L.~P.}\ \bibnamefont {Kouwenhoven}},\ }\href
  {\doibase 10.1126/science.1222360} {\bibfield  {journal} {\bibinfo  {journal}
  {Science}\ }\textbf {\bibinfo {volume} {336}},\ \bibinfo {pages} {1003}
  (\bibinfo {year} {2012})}\BibitemShut {NoStop}%
\bibitem [{\citenamefont {Deng}\ \emph {et~al.}(2012)\citenamefont {Deng},
  \citenamefont {Yu}, \citenamefont {Huang}, \citenamefont {Larsson},
  \citenamefont {Caroff},\ and\ \citenamefont {Xu}}]{Deng2012}%
  \BibitemOpen
  \bibfield  {author} {\bibinfo {author} {\bibfnamefont {M.~T.}\ \bibnamefont
  {Deng}}, \bibinfo {author} {\bibfnamefont {C.~L.}\ \bibnamefont {Yu}},
  \bibinfo {author} {\bibfnamefont {G.~Y.}\ \bibnamefont {Huang}}, \bibinfo
  {author} {\bibfnamefont {M.}~\bibnamefont {Larsson}}, \bibinfo {author}
  {\bibfnamefont {P.}~\bibnamefont {Caroff}}, \ and\ \bibinfo {author}
  {\bibfnamefont {H.~Q.}\ \bibnamefont {Xu}},\ }\href {\doibase
  10.1021/nl303758w} {\bibfield  {journal} {\bibinfo  {journal} {Nano Lett.}\
  }\textbf {\bibinfo {volume} {12}},\ \bibinfo {pages} {6414} (\bibinfo {year}
  {2012})}\BibitemShut {NoStop}%
\bibitem [{\citenamefont {Churchill}\ \emph {et~al.}(2013)\citenamefont
  {Churchill}, \citenamefont {Fatemi}, \citenamefont {Grove-Rasmussen},
  \citenamefont {Deng}, \citenamefont {Caroff}, \citenamefont {Xu},\ and\
  \citenamefont {Marcus}}]{Churchill2013}%
  \BibitemOpen
  \bibfield  {author} {\bibinfo {author} {\bibfnamefont {H.~O.~H.}\
  \bibnamefont {Churchill}}, \bibinfo {author} {\bibfnamefont {V.}~\bibnamefont
  {Fatemi}}, \bibinfo {author} {\bibfnamefont {K.}~\bibnamefont
  {Grove-Rasmussen}}, \bibinfo {author} {\bibfnamefont {M.~T.}\ \bibnamefont
  {Deng}}, \bibinfo {author} {\bibfnamefont {P.}~\bibnamefont {Caroff}},
  \bibinfo {author} {\bibfnamefont {H.~Q.}\ \bibnamefont {Xu}}, \ and\ \bibinfo
  {author} {\bibfnamefont {C.~M.}\ \bibnamefont {Marcus}},\ }\href {\doibase
  10.1103/PhysRevB.87.241401} {\bibfield  {journal} {\bibinfo  {journal}
  {Phys.~Rev.~B}\ }\textbf {\bibinfo {volume} {87}},\ \bibinfo {pages} {241401}
  (\bibinfo {year} {2013})}\BibitemShut {NoStop}%
\bibitem [{\citenamefont {Finck}\ \emph {et~al.}(2013)\citenamefont {Finck},
  \citenamefont {Van~Harlingen}, \citenamefont {Mohseni}, \citenamefont
  {Jung},\ and\ \citenamefont {Li}}]{Finck2013}%
  \BibitemOpen
  \bibfield  {author} {\bibinfo {author} {\bibfnamefont {A.~D.~K.}\
  \bibnamefont {Finck}}, \bibinfo {author} {\bibfnamefont {D.~J.}\ \bibnamefont
  {Van~Harlingen}}, \bibinfo {author} {\bibfnamefont {P.~K.}\ \bibnamefont
  {Mohseni}}, \bibinfo {author} {\bibfnamefont {K.}~\bibnamefont {Jung}}, \
  and\ \bibinfo {author} {\bibfnamefont {X.}~\bibnamefont {Li}},\ }\href
  {\doibase 10.1103/PhysRevLett.110.126406} {\bibfield  {journal} {\bibinfo
  {journal} {Phys. Rev. Lett.}\ }\textbf {\bibinfo {volume} {110}},\ \bibinfo
  {pages} {126406} (\bibinfo {year} {2013})}\BibitemShut {NoStop}%
\bibitem [{\citenamefont {Deng}\ \emph {et~al.}(2016)\citenamefont {Deng},
  \citenamefont {Vaitiek}, \citenamefont {Hansen}, \citenamefont {Danon},
  \citenamefont {Leijnse}, \citenamefont {Flensberg}, \citenamefont
  {Krogstrup},\ and\ \citenamefont {Marcus}}]{Deng2016}%
  \BibitemOpen
  \bibfield  {author} {\bibinfo {author} {\bibfnamefont {M.~T.}\ \bibnamefont
  {Deng}}, \bibinfo {author} {\bibfnamefont {S.}~\bibnamefont {Vaitiek}},
  \bibinfo {author} {\bibfnamefont {E.~B.}\ \bibnamefont {Hansen}}, \bibinfo
  {author} {\bibfnamefont {J.}~\bibnamefont {Danon}}, \bibinfo {author}
  {\bibfnamefont {M.}~\bibnamefont {Leijnse}}, \bibinfo {author} {\bibfnamefont
  {K.}~\bibnamefont {Flensberg}}, \bibinfo {author} {\bibfnamefont
  {P.}~\bibnamefont {Krogstrup}}, \ and\ \bibinfo {author} {\bibfnamefont
  {C.~M.}\ \bibnamefont {Marcus}},\ }\href@noop {} {\bibfield  {journal}
  {\bibinfo  {journal} {Science}\ }\textbf {\bibinfo {volume} {354}},\ \bibinfo
  {pages} {1557} (\bibinfo {year} {2016})}\BibitemShut {NoStop}%
\bibitem [{\citenamefont {Nichele}\ \emph {et~al.}(2017)\citenamefont
  {Nichele}, \citenamefont {Drachmann}, \citenamefont {Whiticar}, \citenamefont
  {O'Farrell}, \citenamefont {Suominen}, \citenamefont {Fornieri},
  \citenamefont {Wang}, \citenamefont {Gardner}, \citenamefont {Thomas},
  \citenamefont {Hatke}, \citenamefont {Krogstrup}, \citenamefont {Manfra},
  \citenamefont {Flensberg},\ and\ \citenamefont {Marcus}}]{Nichele2017}%
  \BibitemOpen
  \bibfield  {author} {\bibinfo {author} {\bibfnamefont {F.}~\bibnamefont
  {Nichele}}, \bibinfo {author} {\bibfnamefont {A.~C.}\ \bibnamefont
  {Drachmann}}, \bibinfo {author} {\bibfnamefont {A.~M.}\ \bibnamefont
  {Whiticar}}, \bibinfo {author} {\bibfnamefont {E.~C.}\ \bibnamefont
  {O'Farrell}}, \bibinfo {author} {\bibfnamefont {H.~J.}\ \bibnamefont
  {Suominen}}, \bibinfo {author} {\bibfnamefont {A.}~\bibnamefont {Fornieri}},
  \bibinfo {author} {\bibfnamefont {T.}~\bibnamefont {Wang}}, \bibinfo {author}
  {\bibfnamefont {G.~C.}\ \bibnamefont {Gardner}}, \bibinfo {author}
  {\bibfnamefont {C.}~\bibnamefont {Thomas}}, \bibinfo {author} {\bibfnamefont
  {A.~T.}\ \bibnamefont {Hatke}}, \bibinfo {author} {\bibfnamefont
  {P.}~\bibnamefont {Krogstrup}}, \bibinfo {author} {\bibfnamefont {M.~J.}\
  \bibnamefont {Manfra}}, \bibinfo {author} {\bibfnamefont {K.}~\bibnamefont
  {Flensberg}}, \ and\ \bibinfo {author} {\bibfnamefont {C.~M.}\ \bibnamefont
  {Marcus}},\ }\href {\doibase 10.1103/PhysRevLett.119.136803} {\bibfield
  {journal} {\bibinfo  {journal} {Phys.~Rev.~Lett.}\ }\textbf {\bibinfo
  {volume} {119}},\ \bibinfo {pages} {136803} (\bibinfo {year}
  {2017})}\BibitemShut {NoStop}%
\bibitem [{\citenamefont {Zhang}\ \emph {et~al.}(2018)\citenamefont {Zhang},
  \citenamefont {Liu}, \citenamefont {Gazibegovic}, \citenamefont {Xu},
  \citenamefont {Logan}, \citenamefont {Wang}, \citenamefont {van Loo},
  \citenamefont {Bommer}, \citenamefont {de~Moor}, \citenamefont {Car},
  \citenamefont {het Veld}, \citenamefont {van Veldhoven}, \citenamefont
  {Koelling}, \citenamefont {Verheijen}, \citenamefont {Pendharkar},
  \citenamefont {Pennachio}, \citenamefont {Shojaei}, \citenamefont {Lee},
  \citenamefont {Palmstrom}, \citenamefont {Bakkers}, \citenamefont {Sarma},\
  and\ \citenamefont {Kouwenhoven}}]{Zhang2018}%
  \BibitemOpen
  \bibfield  {author} {\bibinfo {author} {\bibfnamefont {H.}~\bibnamefont
  {Zhang}}, \bibinfo {author} {\bibfnamefont {C.-X.}\ \bibnamefont {Liu}},
  \bibinfo {author} {\bibfnamefont {S.}~\bibnamefont {Gazibegovic}}, \bibinfo
  {author} {\bibfnamefont {D.}~\bibnamefont {Xu}}, \bibinfo {author}
  {\bibfnamefont {J.~A.}\ \bibnamefont {Logan}}, \bibinfo {author}
  {\bibfnamefont {G.}~\bibnamefont {Wang}}, \bibinfo {author} {\bibfnamefont
  {N.}~\bibnamefont {van Loo}}, \bibinfo {author} {\bibfnamefont {J.~D.~S.}\
  \bibnamefont {Bommer}}, \bibinfo {author} {\bibfnamefont {M.~W.~A.}\
  \bibnamefont {de~Moor}}, \bibinfo {author} {\bibfnamefont {D.}~\bibnamefont
  {Car}}, \bibinfo {author} {\bibfnamefont {R.~L. M.~O.}\ \bibnamefont {het
  Veld}}, \bibinfo {author} {\bibfnamefont {P.~J.}\ \bibnamefont {van
  Veldhoven}}, \bibinfo {author} {\bibfnamefont {S.}~\bibnamefont {Koelling}},
  \bibinfo {author} {\bibfnamefont {M.~A.}\ \bibnamefont {Verheijen}}, \bibinfo
  {author} {\bibfnamefont {M.}~\bibnamefont {Pendharkar}}, \bibinfo {author}
  {\bibfnamefont {D.~J.}\ \bibnamefont {Pennachio}}, \bibinfo {author}
  {\bibfnamefont {B.}~\bibnamefont {Shojaei}}, \bibinfo {author} {\bibfnamefont
  {J.~S.}\ \bibnamefont {Lee}}, \bibinfo {author} {\bibfnamefont {C.~J.}\
  \bibnamefont {Palmstrom}}, \bibinfo {author} {\bibfnamefont {E.~P. A.~M.}\
  \bibnamefont {Bakkers}}, \bibinfo {author} {\bibfnamefont {S.~D.}\
  \bibnamefont {Sarma}}, \ and\ \bibinfo {author} {\bibfnamefont {L.~P.}\
  \bibnamefont {Kouwenhoven}},\ }\href {\doibase 10.1038/nature26142}
  {\bibfield  {journal} {\bibinfo  {journal} {Nature}\ }\textbf {\bibinfo
  {volume} {556}},\ \bibinfo {pages} {74} (\bibinfo {year} {2018})}\BibitemShut
  {NoStop}%
\bibitem [{Den()}]{Deng2018}%
  \BibitemOpen
  \href@noop {} {}\bibinfo {note} {M. T. Deng, S. Vaitiekenas, E. Prada, P.
  San-Jose, J. Nyg\r{a}rd, P. Krogstrup, R. Aguado, C. M. Marcus,
  \href{https://arxiv.org/abs/1712.03536}{arXiv:1712.03536}}\BibitemShut
  {NoStop}%
\bibitem [{\citenamefont {G\"{u}l}\ \emph {et~al.}(2018)\citenamefont
  {G\"{u}l}, \citenamefont {Zhang}, \citenamefont {Bommer}, \citenamefont
  {de~Moor}, \citenamefont {Car}, \citenamefont {Plissard}, \citenamefont
  {Bakkers}, \citenamefont {Geresdi}, \citenamefont {Watanabe}, \citenamefont
  {Taniguchi},\ and\ \citenamefont {Kouwenhoven}}]{Gul2018}%
  \BibitemOpen
  \bibfield  {author} {\bibinfo {author} {\bibfnamefont {O.}~\bibnamefont
  {G\"{u}l}}, \bibinfo {author} {\bibfnamefont {H.}~\bibnamefont {Zhang}},
  \bibinfo {author} {\bibfnamefont {J.~D.~S.}\ \bibnamefont {Bommer}}, \bibinfo
  {author} {\bibfnamefont {M.~W.~A.}\ \bibnamefont {de~Moor}}, \bibinfo
  {author} {\bibfnamefont {D.}~\bibnamefont {Car}}, \bibinfo {author}
  {\bibfnamefont {S.~R.}\ \bibnamefont {Plissard}}, \bibinfo {author}
  {\bibfnamefont {E.~P. A.~M.}\ \bibnamefont {Bakkers}}, \bibinfo {author}
  {\bibfnamefont {A.}~\bibnamefont {Geresdi}}, \bibinfo {author} {\bibfnamefont
  {K.}~\bibnamefont {Watanabe}}, \bibinfo {author} {\bibfnamefont
  {T.}~\bibnamefont {Taniguchi}}, \ and\ \bibinfo {author} {\bibfnamefont
  {L.~P.}\ \bibnamefont {Kouwenhoven}},\ }\href {\doibase
  10.1038/s41565-017-0032-8} {\bibfield  {journal} {\bibinfo  {journal} {Nature
  Nanotechnology}\ }\textbf {\bibinfo {volume} {13}},\ \bibinfo {pages} {192}
  (\bibinfo {year} {2018})}\BibitemShut {NoStop}%
\bibitem [{\citenamefont {Ivanov}(2001)}]{Ivanov2001}%
  \BibitemOpen
  \bibfield  {author} {\bibinfo {author} {\bibfnamefont {D.~A.}\ \bibnamefont
  {Ivanov}},\ }\href {\doibase 10.1103/PhysRevLett.86.268} {\bibfield
  {journal} {\bibinfo  {journal} {Phys. Rev. Lett.}\ }\textbf {\bibinfo
  {volume} {86}},\ \bibinfo {pages} {268} (\bibinfo {year} {2001})}\BibitemShut
  {NoStop}%
\bibitem [{\citenamefont {Alicea}\ \emph {et~al.}(2011)\citenamefont {Alicea},
  \citenamefont {Oreg}, \citenamefont {Refael}, \citenamefont {von Oppen},\
  and\ \citenamefont {Fisher}}]{Alicea2011}%
  \BibitemOpen
  \bibfield  {author} {\bibinfo {author} {\bibfnamefont {J.}~\bibnamefont
  {Alicea}}, \bibinfo {author} {\bibfnamefont {Y.}~\bibnamefont {Oreg}},
  \bibinfo {author} {\bibfnamefont {G.}~\bibnamefont {Refael}}, \bibinfo
  {author} {\bibfnamefont {F.}~\bibnamefont {von Oppen}}, \ and\ \bibinfo
  {author} {\bibfnamefont {M.~P.~A.}\ \bibnamefont {Fisher}},\ }\href {\doibase
  10.1038/nphys1915} {\bibfield  {journal} {\bibinfo  {journal} {Nat.~Phys.}\
  }\textbf {\bibinfo {volume} {7}},\ \bibinfo {pages} {412} (\bibinfo {year}
  {2011})}\BibitemShut {NoStop}%
\bibitem [{\citenamefont {Flensberg}(2011)}]{Flensberg2011}%
  \BibitemOpen
  \bibfield  {author} {\bibinfo {author} {\bibfnamefont {K.}~\bibnamefont
  {Flensberg}},\ }\href {\doibase 10.1103/PhysRevLett.106.090503} {\bibfield
  {journal} {\bibinfo  {journal} {Phys.~Rev.~Lett.}\ }\textbf {\bibinfo
  {volume} {106}},\ \bibinfo {pages} {090503} (\bibinfo {year}
  {2011})}\BibitemShut {NoStop}%
\bibitem [{\citenamefont {Hyart}\ \emph {et~al.}(2013)\citenamefont {Hyart},
  \citenamefont {Heck}, \citenamefont {Fulga}, \citenamefont {Burello},
  \citenamefont {Akhmerov},\ and\ \citenamefont {Beenakker}}]{Hyart2013}%
  \BibitemOpen
  \bibfield  {author} {\bibinfo {author} {\bibfnamefont {T.}~\bibnamefont
  {Hyart}}, \bibinfo {author} {\bibfnamefont {B.~V.}\ \bibnamefont {Heck}},
  \bibinfo {author} {\bibfnamefont {I.}~\bibnamefont {Fulga}}, \bibinfo
  {author} {\bibfnamefont {M.}~\bibnamefont {Burello}}, \bibinfo {author}
  {\bibfnamefont {A.~R.}\ \bibnamefont {Akhmerov}}, \ and\ \bibinfo {author}
  {\bibfnamefont {C.~W.~J.}\ \bibnamefont {Beenakker}},\ }\href {\doibase
  10.1103/PhysRevB.88.035121} {\bibfield  {journal} {\bibinfo  {journal}
  {Phys.~Rev.~B}\ }\textbf {\bibinfo {volume} {88}},\ \bibinfo {pages} {035121}
  (\bibinfo {year} {2013})}\BibitemShut {NoStop}%
\bibitem [{\citenamefont {Aasen}\ \emph {et~al.}(2016)\citenamefont {Aasen},
  \citenamefont {Hell}, \citenamefont {Mishmash}, \citenamefont {Higginbotham},
  \citenamefont {Danon}, \citenamefont {Leijnse}, \citenamefont {Jespersen},
  \citenamefont {Folk}, \citenamefont {Marcus}, \citenamefont {Flensberg},\
  and\ \citenamefont {Alicea}}]{Aasen2016}%
  \BibitemOpen
  \bibfield  {author} {\bibinfo {author} {\bibfnamefont {D.}~\bibnamefont
  {Aasen}}, \bibinfo {author} {\bibfnamefont {M.}~\bibnamefont {Hell}},
  \bibinfo {author} {\bibfnamefont {R.~V.}\ \bibnamefont {Mishmash}}, \bibinfo
  {author} {\bibfnamefont {A.}~\bibnamefont {Higginbotham}}, \bibinfo {author}
  {\bibfnamefont {J.}~\bibnamefont {Danon}}, \bibinfo {author} {\bibfnamefont
  {M.}~\bibnamefont {Leijnse}}, \bibinfo {author} {\bibfnamefont {T.~S.}\
  \bibnamefont {Jespersen}}, \bibinfo {author} {\bibfnamefont {J.~A.}\
  \bibnamefont {Folk}}, \bibinfo {author} {\bibfnamefont {C.~M.}\ \bibnamefont
  {Marcus}}, \bibinfo {author} {\bibfnamefont {K.}~\bibnamefont {Flensberg}}, \
  and\ \bibinfo {author} {\bibfnamefont {J.}~\bibnamefont {Alicea}},\ }\href
  {\doibase 10.1103/PhysRevX.6.031016} {\bibfield  {journal} {\bibinfo
  {journal} {Phys.~Rev.~X}\ }\textbf {\bibinfo {volume} {6}},\ \bibinfo {pages}
  {031016} (\bibinfo {year} {2016})}\BibitemShut {NoStop}%
\bibitem [{\citenamefont {Hell}\ \emph {et~al.}(2016)\citenamefont {Hell},
  \citenamefont {Danon}, \citenamefont {Flensberg},\ and\ \citenamefont
  {Leijnse}}]{Hell2016}%
  \BibitemOpen
  \bibfield  {author} {\bibinfo {author} {\bibfnamefont {M.}~\bibnamefont
  {Hell}}, \bibinfo {author} {\bibfnamefont {J.}~\bibnamefont {Danon}},
  \bibinfo {author} {\bibfnamefont {K.}~\bibnamefont {Flensberg}}, \ and\
  \bibinfo {author} {\bibfnamefont {M.}~\bibnamefont {Leijnse}},\ }\href@noop
  {} {\bibfield  {journal} {\bibinfo  {journal} {Phys.~Rev.~B}\ }\textbf
  {\bibinfo {volume} {94}},\ \bibinfo {pages} {035424} (\bibinfo {year}
  {2016})}\BibitemShut {NoStop}%
\bibitem [{\citenamefont {Landau}\ \emph {et~al.}(2016)\citenamefont {Landau},
  \citenamefont {Plugge}, \citenamefont {Sela}, \citenamefont {Altland},
  \citenamefont {Albrecht},\ and\ \citenamefont {Egger}}]{Landau2016}%
  \BibitemOpen
  \bibfield  {author} {\bibinfo {author} {\bibfnamefont {L.~A.}\ \bibnamefont
  {Landau}}, \bibinfo {author} {\bibfnamefont {S.}~\bibnamefont {Plugge}},
  \bibinfo {author} {\bibfnamefont {E.}~\bibnamefont {Sela}}, \bibinfo {author}
  {\bibfnamefont {A.}~\bibnamefont {Altland}}, \bibinfo {author} {\bibfnamefont
  {S.~M.}\ \bibnamefont {Albrecht}}, \ and\ \bibinfo {author} {\bibfnamefont
  {R.}~\bibnamefont {Egger}},\ }\href {\doibase 10.1103/PhysRevLett.116.050501}
  {\bibfield  {journal} {\bibinfo  {journal} {Phys.~Rev.~Lett.}\ }\textbf
  {\bibinfo {volume} {116}},\ \bibinfo {pages} {050501} (\bibinfo {year}
  {2016})}\BibitemShut {NoStop}%
\bibitem [{\citenamefont {Plugge}\ \emph {et~al.}()\citenamefont {Plugge},
  \citenamefont {Rasmussen}, \citenamefont {Egger},\ and\ \citenamefont
  {Flensberg}}]{Plugge2017}%
  \BibitemOpen
  \bibfield  {author} {\bibinfo {author} {\bibfnamefont {S.}~\bibnamefont
  {Plugge}}, \bibinfo {author} {\bibfnamefont {A.}~\bibnamefont {Rasmussen}},
  \bibinfo {author} {\bibfnamefont {R.}~\bibnamefont {Egger}}, \ and\ \bibinfo
  {author} {\bibfnamefont {K.}~\bibnamefont {Flensberg}},\ }\href {\doibase
  10.1088/1367-2630/aa54e1} {\bibfield  {journal} {\bibinfo  {journal} {New
  Journal of Physics}\ }\textbf {\bibinfo {volume} {19}},\ \bibinfo {pages}
  {012001}}\BibitemShut {NoStop}%
\bibitem [{\citenamefont {Karzig}\ \emph {et~al.}(2017)\citenamefont {Karzig},
  \citenamefont {Knapp}, \citenamefont {Lutchyn}, \citenamefont {Bonderson},
  \citenamefont {Hastings}, \citenamefont {Nayak}, \citenamefont {Alicea},
  \citenamefont {Flensberg}, \citenamefont {Plugge}, \citenamefont {Oreg},
  \citenamefont {Marcus},\ and\ \citenamefont {Freedman}}]{Karzig2017}%
  \BibitemOpen
  \bibfield  {author} {\bibinfo {author} {\bibfnamefont {T.}~\bibnamefont
  {Karzig}}, \bibinfo {author} {\bibfnamefont {C.}~\bibnamefont {Knapp}},
  \bibinfo {author} {\bibfnamefont {R.~M.}\ \bibnamefont {Lutchyn}}, \bibinfo
  {author} {\bibfnamefont {P.}~\bibnamefont {Bonderson}}, \bibinfo {author}
  {\bibfnamefont {M.~B.}\ \bibnamefont {Hastings}}, \bibinfo {author}
  {\bibfnamefont {C.}~\bibnamefont {Nayak}}, \bibinfo {author} {\bibfnamefont
  {J.}~\bibnamefont {Alicea}}, \bibinfo {author} {\bibfnamefont
  {K.}~\bibnamefont {Flensberg}}, \bibinfo {author} {\bibfnamefont
  {S.}~\bibnamefont {Plugge}}, \bibinfo {author} {\bibfnamefont
  {Y.}~\bibnamefont {Oreg}}, \bibinfo {author} {\bibfnamefont {C.~M.}\
  \bibnamefont {Marcus}}, \ and\ \bibinfo {author} {\bibfnamefont {M.~H.}\
  \bibnamefont {Freedman}},\ }\href {\doibase 10.1103/PhysRevB.95.235305}
  {\bibfield  {journal} {\bibinfo  {journal} {Phys.~Rev.~B}\ }\textbf {\bibinfo
  {volume} {95}},\ \bibinfo {pages} {235305} (\bibinfo {year}
  {2017})}\BibitemShut {NoStop}%
\bibitem [{Sch()}]{Schrade2018}%
  \BibitemOpen
  \href@noop {} {}\bibinfo {note} {C. Schrade and L. Fu,
  \href{https://arxiv.org/abs/1803.01002}{arXiv:1803.01002}}\BibitemShut
  {NoStop}%
\bibitem [{\citenamefont {Ginossar}\ and\ \citenamefont
  {Grosfeld}(2014)}]{Ginossar2014}%
  \BibitemOpen
  \bibfield  {author} {\bibinfo {author} {\bibfnamefont {E.}~\bibnamefont
  {Ginossar}}\ and\ \bibinfo {author} {\bibfnamefont {E.}~\bibnamefont
  {Grosfeld}},\ }\href {\doibase 10.1038/ncomms5772} {\bibfield  {journal}
  {\bibinfo  {journal} {Nature Communications}\ }\textbf {\bibinfo {volume}
  {5}},\ \bibinfo {pages} {4772} (\bibinfo {year} {2014})}\BibitemShut
  {NoStop}%
\bibitem [{\citenamefont {Yavilberg}\ \emph {et~al.}(2015)\citenamefont
  {Yavilberg}, \citenamefont {Ginossar},\ and\ \citenamefont
  {Grosfeld}}]{Yavilberg2015}%
  \BibitemOpen
  \bibfield  {author} {\bibinfo {author} {\bibfnamefont {K.}~\bibnamefont
  {Yavilberg}}, \bibinfo {author} {\bibfnamefont {E.}~\bibnamefont {Ginossar}},
  \ and\ \bibinfo {author} {\bibfnamefont {E.}~\bibnamefont {Grosfeld}},\
  }\href {\doibase 10.1103/PhysRevB.92.075143} {\bibfield  {journal} {\bibinfo
  {journal} {Phys.~Rev.~B}\ }\textbf {\bibinfo {volume} {92}},\ \bibinfo
  {pages} {075143} (\bibinfo {year} {2015})},\ \Eprint
  {http://arxiv.org/abs/1411.5699} {1411.5699} \BibitemShut {NoStop}%
\bibitem [{\citenamefont {Koch}\ \emph {et~al.}(2007)\citenamefont {Koch},
  \citenamefont {Yu}, \citenamefont {Gambetta}, \citenamefont {Houck},
  \citenamefont {Schuster}, \citenamefont {Majer}, \citenamefont {Blais},
  \citenamefont {Devoret}, \citenamefont {Girvin},\ and\ \citenamefont
  {Schoelkopf}}]{Koch2007}%
  \BibitemOpen
  \bibfield  {author} {\bibinfo {author} {\bibfnamefont {J.}~\bibnamefont
  {Koch}}, \bibinfo {author} {\bibfnamefont {T.~M.}\ \bibnamefont {Yu}},
  \bibinfo {author} {\bibfnamefont {J.}~\bibnamefont {Gambetta}}, \bibinfo
  {author} {\bibfnamefont {A.~A.}\ \bibnamefont {Houck}}, \bibinfo {author}
  {\bibfnamefont {D.~I.}\ \bibnamefont {Schuster}}, \bibinfo {author}
  {\bibfnamefont {J.}~\bibnamefont {Majer}}, \bibinfo {author} {\bibfnamefont
  {A.}~\bibnamefont {Blais}}, \bibinfo {author} {\bibfnamefont {M.~H.}\
  \bibnamefont {Devoret}}, \bibinfo {author} {\bibfnamefont {S.~M.}\
  \bibnamefont {Girvin}}, \ and\ \bibinfo {author} {\bibfnamefont {R.~J.}\
  \bibnamefont {Schoelkopf}},\ }\href {\doibase 10.1103/PhysRevA.76.042319}
  {\bibfield  {journal} {\bibinfo  {journal} {Phys. Rev. A}\ }\textbf {\bibinfo
  {volume} {76}},\ \bibinfo {pages} {042319} (\bibinfo {year}
  {2007})}\BibitemShut {NoStop}%
\bibitem [{\citenamefont {Petta}\ \emph {et~al.}(2005)\citenamefont {Petta},
  \citenamefont {Johnson}, \citenamefont {Taylor}, \citenamefont {Laird},
  \citenamefont {Yacoby}, \citenamefont {Lukin}, \citenamefont {Marcus},
  \citenamefont {Hanson},\ and\ \citenamefont {Gossard}}]{Petta2005}%
  \BibitemOpen
  \bibfield  {author} {\bibinfo {author} {\bibfnamefont {J.~R.}\ \bibnamefont
  {Petta}}, \bibinfo {author} {\bibfnamefont {A.~C.}\ \bibnamefont {Johnson}},
  \bibinfo {author} {\bibfnamefont {J.~M.}\ \bibnamefont {Taylor}}, \bibinfo
  {author} {\bibfnamefont {E.~A.}\ \bibnamefont {Laird}}, \bibinfo {author}
  {\bibfnamefont {A.}~\bibnamefont {Yacoby}}, \bibinfo {author} {\bibfnamefont
  {M.~D.}\ \bibnamefont {Lukin}}, \bibinfo {author} {\bibfnamefont {C.~M.}\
  \bibnamefont {Marcus}}, \bibinfo {author} {\bibfnamefont {M.~P.}\
  \bibnamefont {Hanson}}, \ and\ \bibinfo {author} {\bibfnamefont {A.~C.}\
  \bibnamefont {Gossard}},\ }\href@noop {} {\bibfield  {journal} {\bibinfo
  {journal} {Science}\ }\textbf {\bibinfo {volume} {309}},\ \bibinfo {pages}
  {2180} (\bibinfo {year} {2005})}\BibitemShut {NoStop}%
\end{thebibliography}%
\bibliographystyle{apsrev4-1}

\end{document}